\documentclass[fleqn,usenatbib]{mnras}

\usepackage[T1]{fontenc}
\usepackage{ae,aecompl}

\usepackage{graphicx}	
\usepackage{amsmath}	
\usepackage{amssymb}	
\usepackage{threeparttable}
\usepackage{multirow}
\usepackage{url}
\usepackage{multicol}
\usepackage{xcolor}

\newcommand{\mum}{\ifmmode{\rm \mu m}\else{$\mu$m}\fi}

\newcommand{\Msun}{\ensuremath{{\rm M}_{\odot}}}   
  
\newcommand{\chisq}{\ifmmode{\chi^{2} }\else{$\chi^2$}\fi}
\newcommand{\rchisq}{\ifmmode{\chi^{2} }\else{$\chi^2_\nu$}\fi}

\newcommand{\starbug}{\textsc{starbugii}}

\usepackage{orcidlink}
\usepackage{academicons}
\definecolor{orcidlogocol}{HTML}{A6CE39}

\usepackage{newtxtext,newtxmath}

\title[JWST/MIRI Star Formation in Cen A]{MICONIC: The spatial relationship between star formation and the AGN in Centaurus A revealed by JWST/MIRI}

\author
[O.\ C.\ Jones et al.]{O.\ C.\ Jones$^{1}$\thanks{E-mail: olivia.jones@stfc.ac.uk}\orcidlink{0000-0003-4870-5547},
M.\ Jones$^{1}$\orcidlink{0009-0003-3542-6896},
D.\ Dicken$^{1}$\orcidlink{0000-0003-0589-5969},
G.\ S.\ Wright$^{1}$\orcidlink{0000-0001-7416-7936},
M.\ Garc\'{\i}a Mar\'{\i}n$^{2}$\orcidlink{0000-0003-4801-0489},
\newauthor
A.\ Alonso Herrero$^{3}$\orcidlink{0000-0001-6794-2519},
P.\ Guillard$^{4,5}$\orcidlink{0000-0002-2421-1350},
K.\ Justtanont$^{6}$\orcidlink{0000-0003-1689-9201},
M.\ Meixner$^{7}$\orcidlink{0000-0002-0522-3743},
A.\ Labiano$^{8}$\orcidlink{0000-0002-0690-8824},
\newauthor
D.\ Rouan$^{9}$\orcidlink{0000-0002-2352-1736},
P.\ van der Werf$^{10}$\orcidlink{0000-0001-5434-5942},
L.\ Pantoni$^{11}$\orcidlink{0000-0003-2666-5759},
V.\ A.\ Buiten$^{10}$\orcidlink{0009-0003-4835-2435},
T.\ B\"oker$^{12}$\orcidlink{0000-0002-5666-7782},
\newauthor
G.\ {\"O}stlin$^{13}$\orcidlink{0000-0002-3005-1349}, 
L.\ Evangelista$^{4}$\orcidlink{0009-0000-6990-7928},
M.\ Baes$^{11}$\orcidlink{0000-0002-3930-2757},
L.\ Colina$^{14}$\orcidlink{0000-0002-9090-4227},
L.\ Hermosa Mu{\~n}oz$^{15}$\orcidlink{0000-0002-9610-0123},
\newauthor
Th.\ Henning$^{16}$\orcidlink{0000-0002-1493-300X},
M.\ G\"udel$^{17,18,19}$\orcidlink{0000-0001-9818-0588},
T.\ P.\ Ray$^{20}$\orcidlink{0000-0002-2110-1068},
P.-O.\ Lagage$^{21}$
\\
$^{1}$ UK Astronomy Technology Centre, Royal Observatory, Blackford Hill, Edinburgh, EH9 3HJ, UK \\
$^{2}$ Space Telescope Science Institute, 3700 San Martin Drive, Baltimore, MD 21218, USA, \\ 
$^{3}$ Centro de Astrobiolog\'{\i}a (CAB), CSIC--INTA, Camino Bajo del Castillo s/n,
E-28692 Villanueva de la Ca\~nada, Madrid, Spain \\
$^{4}$ Sorbonne Universit\'e, CNRS, UMR 7095, Institut d’Astrophysique de Paris, 98bis bd Arago, 75014 Paris, France \\
$^{5}$ Institut Universitaire de France, Minist\'ere de l’Enseignement Supérieur et de la Recherche, 1 rue Descartes, 75231 Paris Cedex 05, France \\
$^{6}$ Department of Physics and Astronomy, Chalmers University of Technology, 412 96 Gothenburg, Sweden \\
$^{7}$ Jet Propulsion Laboratory, California Institute of Technology, 4800 Oak Grove Dr., Pasadena, CA 91109, USA \\
$^{8}$ Telespazio UK for the European Space Agency (ESA), ESAC, Camino Bajo del Castillo s/n, E-28692 Villanueva de la Ca\~nada, Madrid, Spain\\
$^{9}$ LIRA, Observatoire de Paris, Universit\'e PSL, Sorbonne Universit\'e, Universit\'e Paris Cit\'e, CY Cergy Paris Universit\'e, CNRS, 92190 Meudon, France \\
$^{10}$ Leiden Observatory, Leiden University, PO Box 9513 2300 RA Leiden, The Netherlands  \\
$^{11}$ Department of Physics and Astronomy, Universiteit Gent, Proeftuinstraat 86 N3, B-9000 Ghent, Belgium \\
$^{12}$ European Space Agency, c/o STScI, 3700 San Martin Drive, Baltimore, MD 21218, USA\\
$^{13}$ The Oskar Klein Centre, Department of Astronomy, Stockholm University, AlbaNova, SE-10691 Stockholm, Sweden\\ 
$^{14}$ Centro de Astrobiolog\'{\i}a (CAB), CSIC-INTA, Ctra. de Ajalvir km 4, Torrej\'on de Ardoz, E-28850, Madrid, Spain \\
$^{15}$ Departamento de F{\'i}sica, Universidad de Oviedo, Campus de Llamaquique, C/ Calvo Sotelo s/n, 33007 Oviedo, Spain \\
$^{16}$ Max-Planck-Institut f\"ur Astronomie (MPIA), K\"onigstuhl 17, D-69117 Heidelberg, Germany \\
$^{17}$ Dept. of Astrophysics, University of Vienna, T\"urkenschanzstr. 17, A-1180 Vienna, Austria \\
$^{18}$ ETH Z\"urich, Institute for Particle Physics and Astrophysics, Wolfgang-Pauli-Str. 27, 8093 Z\"urich, Switzerland \\
$^{19}$ ASTRON, Netherlands Institute for Radio Astronomy, Oude Hoogeveensedijk 4, 7991 PD Dwingeloo, The Netherlands \\
$^{20}$ School of Cosmic Physics, Dublin Institute for Advanced Studies, 31 Fitzwilliam Place, Dublin, D02 XF86, Ireland \\
$^{21}$ Universit\'e Paris-Saclay, Universit\'e Paris\textbf{} Cit\'e, CEA, CNRS, AIM, F-91191 Gif-sur-Yvette, France
}

\date{Accepted XXX. Received YYY; in original form ZZZ}

\pubyear{2026}

\begin{document}
\label{firstpage}
\pagerange{\pageref{firstpage}--\pageref{lastpage}}
\maketitle


\begin{abstract}
Centaurus~A (Cen~A), the nearest active radio galaxy, hosts a warped dust disc formed in a gas-rich merger. We present \emph{JWST}/MIRI imaging in three filters, F560W, F770W, and F1130W, of this central disc over a $\sim4\times2$ kpc region to characterise its resolved mid-infrared stellar populations. 
The images reveal a system of extended dusty structures, previously identified with \textit{Spitzer} as an ``oval dusty shell'', now resolved into multiple loop-like features that are brightest in F1130W and closely associated with the warped disc.
Colour–magnitude and colour–colour diagnostics reveal a distinct population of 928 red point sources with strong infrared excess, accounting for $\sim$36 per cent of sources with high-quality photometry in all three bands, spatially confined to the disc. These sources exhibit rising mid-infrared spectral slopes indicative of emission from warm dust. Their colours and spatial distribution are consistent with a population dominated by embedded young stellar objects, tracing recent ($\sim10^{5}$–$10^{6}$~yr) star formation within the disc. 
The strong geometric alignment of these sources with the disc, together with the lack of correlation with the radio jet, suggests that star formation in the central regions of Cen~A is primarily regulated by merger-accreted gas, with no strong evidence for AGN jet--ISM interactions.
\end{abstract}





\section{Introduction}\label{intro}

Centaurus~A (hereafter Cen~A; NGC~5128) is the nearest active radio galaxy, at a distance of $D \simeq 3.8$\,Mpc \citep{Neumayer2007, Harris2010}. It is a giant elliptical galaxy hosting an active galactic nucleus (AGN; $L_{\rm bol} \sim 10^{43}$ -- $4\times10^{43}$\,erg\,s$^{-1}$) and a prominent, complex dust lane crossing its nucleus, interpreted as the remnant of a merger with a gas-rich disc galaxy \citep[e.g.][]{Israel1998,Quillen2006}. Cen~A also drives a powerful radio jet extending over hundreds of kiloparsecs, oriented approximately perpendicular to the circumnuclear warped disc \citep{Clarke1992, Hardcastle2003, Espada2017}. These characteristics make it a key laboratory for studying the interplay between AGN activity and the interstellar medium (ISM).

The merger is estimated to have occurred $\sim$2\,Gyr ago \citep{Wang2020}, although its present-day structure may also reflect earlier assembly and subsequent growth through minor mergers and accretion \citep[e.g.,][]{Rejkuba2022}. 
The prominent dust lane and associated shells trace a warped disc extending $\sim$12\,kpc across the galaxy, with a median position angle of $122^\circ \pm 4^\circ$ and a near edge-on inclination \citep{Graham1979, Quillen2010}. 
The accreted disc hosts star formation at a rate of $\sim 0.1$--$1\,M_\odot\,\mathrm{yr}^{-1}$ \citep{Marconi2000}. This episode likely began $\sim$50\,Myr ago and is traced by H\,{\sc ii} regions, OB associations, and infrared recombination-line emission (e.g. Pa$\alpha$) along the dust lane \citep{Dufour1979,Hodge1983,Schreier1996,Marconi2000}.

Recent star formation in the central regions is linked to the accretion of gas-rich material, which may have replenished the disc \citep[e.g.][]{Quillen1993}. 
The star-forming disc extends to radii of $\sim$4\,kpc, while neutral hydrogen (H\,{\sc i}) is detected out to $\sim$7\,kpc \citep{Quillen1992, vanGorkom1990, Schiminovich1994}. 
Kinematically, the star-forming disc is distinct from the elliptical component: the stars and globular clusters exhibit low rotation, whereas the gas- and dust-rich disc shows substantially higher rotational velocities \citep{Quillen1992}. Together, these properties indicate a dynamically complex system in which merger-driven accretion has established a long-lived, actively star-forming disc embedded within an early-type galaxy.

On larger scales, Cen~A exhibits an extended stellar halo with shells and streams that trace its hierarchical assembly history, consistent with continued growth through mergers and accretion \citep{Rejkuba2022}.
Resolved stellar population studies have primarily targeted halo fields several kiloparsecs from the galaxy centre, where stellar crowding and extinction by the central dust lane are reduced. 
Deep optical and near-infrared imaging of these outer regions, spanning $\sim$10--40\,kpc ($\sim$1.5--7\,$r_{\rm eff}$), has resolved individual red giant branch (RGB) and asymptotic giant branch (AGB) stars down to the horizontal branch \citep[e.g.][]{Soria1996,Harris2002,Rejkuba2005}.
These studies show that the halo is dominated by an old ($\gtrsim 8$ Gyr), moderately metal-rich stellar population ($-1.0 \lesssim [\mathrm{M/H}] \lesssim 0.0$), with a significant intermediate-age component (age $\sim$2--4\,Gyr), particularly in the inner halo. This component contributes up to $\sim$10--40\% of the stellar mass and indicates extended star formation over several Gyr. While the bulk of the halo stars formed more than $\sim$5--8 Gyr ago, the presence of this intermediate-age population demonstrates that star formation continued to more recent epochs, consistent with past merger activity \citep{Rejkuba2003,Rejkuba2005,Wang2020}.
Localised star formation is also observed in halo filaments, but occurs at low levels and contributes negligibly to the overall stellar mass of the halo \citep{Mould2000,Crockett2012,Keel2019}. 

Disentangling the drivers of star formation in AGN host galaxies remains challenging, as it may arise either from large-scale gas inflows induced by mergers or from localised compression of the ISM through AGN-driven outflows and jets. In Cen~A, very young ($\lesssim 10$--15~Myr) stellar populations have been identified in $\sim$kpc-scale filamentary structures located $\sim8$--15~kpc from the nucleus, associated with radio lobes extending over tens to hundreds of kiloparsecs \citep{Mould2000,Crockett2012}. These regions are widely interpreted as sites of jet--ISM interaction, where the radio jet interacts with cold gas \citep{Oosterloo2005,Santoro2015}. In contrast, the compact ($\lesssim 4$~kpc) central disc is thought to be fuelled by merger-driven gas accretion \citep{Wang2020}, highlighting the coexistence of multiple star formation pathways within a single system.

Stellar populations within the ``parallelogram'' of Cen~A, a $\sim$3~kpc dust structure first identified in \textit{Spitzer}/IRAC imaging and interpreted as the projected morphology of the galaxy's dusty, $\sim$10--12~kpc warped disc \citep{Quillen2006,Quillen2008}, remain poorly constrained due to severe internal extinction ($A_V \sim 7$--15 mag, with typical values of $\sim7$ mag in the dust lane and $\sim14$ mag toward the nucleus; \citealt{Marconi2000}), which strongly attenuates optical and near-infrared emission. Mid-infrared observations are therefore required, as they penetrate dense dust and trace both evolved stars and embedded star formation.
Previous \textit{Spitzer} observations revealed polycyclic aromatic hydrocarbon (PAH) emission and warm dust in Cen~A \citep{Quillen2008}, but lacked the spatial resolution to resolve individual stellar sources within the disc. 
\emph{JWST}/MIRI imaging now enables resolved studies of these populations across the central kiloparsecs of Cen~A, while MIRI/MRS observations of the central $\sim7''\times12''$ ($\sim100\times200$ pc) have already revealed parsec-scale structure in the circumnuclear environment, including warm molecular gas, ionised gas, and PAH emission associated with AGN feedback \citep{AlonsoHerrero2025,Evangelista2026,Pantoni2026}.

In this paper, we present \emph{JWST}/MIRI imaging of the central regions of Cen~A, spanning $\sim4\times2$ kpc, to characterise its resolved mid-infrared stellar populations. The observations, data reduction, and photometry are described in Section~\ref{obs_cal}. In Section~\ref{Results}, we use mid‑infrared colour diagnostics to identify dust‑enshrouded sources and analyse their spatial distribution. The nature of these sources and their association with the warped dust disc of Cen~A are discussed in Section~\ref{Disc}, and our conclusions are summarised in Section~\ref{Summ_Conc}.

\section{Observations and Reduction}\label{obs_cal}

\subsection{Observations}
\label{sec:obs}

We present mid-infrared imaging of the central disc and nucleus of the nearby radio galaxy Centaurus~A obtained with the Mid-Infrared Instrument \citep[MIRI;][]{Rieke2015PASP, Wright2023} aboard the \emph{James Webb Space Telescope} \citep[JWST;][]{Gardner2023, Rigby2023}. These observations are part of the MIRI European Consortium Guaranteed Time Observations programme \emph{Mid-Infrared Characterisation of Nearby Iconic Galaxy Centres} (MICONIC).

MIRI imaging \citep{Dicken2024} from Programmes~1269 and~4529 was combined to produce a contiguous $3\times1$ mosaic of the central region and inner disc of Centaurus~A in the F560W, F770W, and F1130W filters, using the \textsc{FULL} detector array and a four-point cycling dither pattern optimised for extended emission. Programme~1269 was carried out on 2023~March~19 and employed MIRI Medium--Resolution Spectroscopy (MRS) as the prime observing mode, using a position angle between 23 and 345 degrees. MIRI imaging data were obtained simultaneously. The associated MRS observations are presented by \citet{AlonsoHerrero2025}, \citet{Pantoni2026}, and \citet{Evangelista2026}. For the imaging component, each filter was observed with 25 groups per integration, two integrations per exposure, and four dithers, resulting in a total on-source integration time of 1199~s per filter.

Dedicated MIRI imaging to extend the spatial coverage of the Centaurus~A disc was obtained in Programme~4529, executed on 2025~July~05. These observations were arranged as two individual pointings in a non-interruptible sequence, completing the three-tile coverage of the Centaurus~A disc. In this case the observation position angles were not constrained, allowing greater scheduling flexibility; instead, different coordinates were employed for each of the two individual mosaic tiles to optimise coverage of the central region within the available observing time. Each tile was observed with 10 groups per integration and two integrations per exposure, again using four dithers per filter, yielding a total on-source integration time of 233~s per filter per tile. A summary of the mid-IR observations is provided in Table~\ref{tab:miri_obs}.

\begin{figure}
    \centering
    \includegraphics[width=\columnwidth]{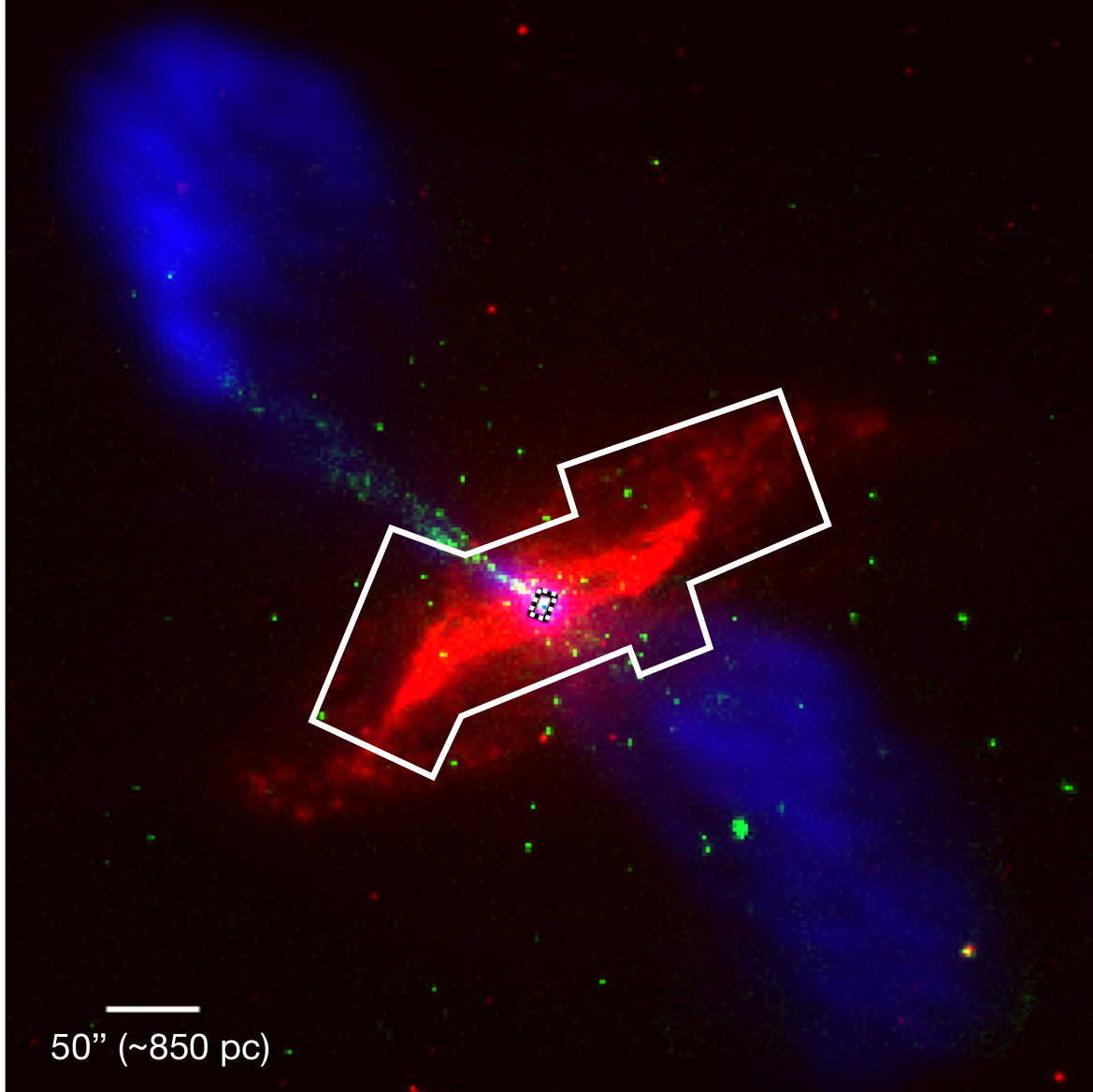}
    \caption{Colour composite image of Centaurus~A: red corresponds to $5.8~\mu\mathrm{m}$ emission from \textit{Spitzer}/IRAC \citep{Quillen2006}, green to X-ray emission from \textit{Chandra}/ACIS \citep{Hardcastle2007_chandra}, and blue to the VLA radio jet \citep{Hardcastle2003}. The MIRI imaging mosaic (this work) is outlined in solid white, while the MICONIC MIRI/MRS footprint of the central $7^{\prime\prime} \times 12^{\prime\prime}$ (see~\citealp{AlonsoHerrero2025, Pantoni2026, Evangelista2026}) is indicated by the black and white dashed line. North is up, and east is to the left.} 
    \label{fig:footprint}
\end{figure}

The final mosaic spans approximately $6.9~\mathrm{arcmin}^2$, corresponding to physical scales of order a few kiloparsecs at the distance of Cen~A ($1'' \simeq 18$\,pc), and encompasses the active nucleus and the surrounding warped dust disc. The footprint of the combined mosaic is shown in Fig.~\ref{fig:footprint}.
It is centred on the nucleus at $\mathrm{RA} = 13^{\mathrm{h}}25^{\mathrm{m}}27.6^{\mathrm{s}}$, $\mathrm{Dec} = -43^{\circ}01^{\prime}09^{\prime\prime}.3$ (J2000). 

\begin{table*}
\caption{Summary of \emph{JWST}/MIRI imaging observations of the disc of Centaurus~A.}
\label{tab:miri_obs}
\centering
\begin{tabular}{lccccccc}
\hline\hline
Programme & Cycle & PI & Date & Filters & Dithers & Groups$\times$Ints & $t_{\mathrm{exp}}$ (s) \\
\hline
1269 & 1 & Luetzgendorf & 2023 03 19 & F560W, F770W, F1130W & 4 & 25$\times$2 & 1199 \\
4529 & 3 & Dicken & 2025 07 05 & F560W, F770W, F1130W & 4 & 10$\times$2 & 233 \\
\hline
\end{tabular}
\begin{flushleft}
\footnotesize
Notes.~All observations used a four-point cycling dither pattern optimised for extended emission and were obtained with the \textsc{FULL} detector array.
Both programmes are part of the MICONIC Guaranteed Time Observations programme.
\end{flushleft}
\end{table*}

\subsection{MIRI Image Processing}
\label{sec:Reduction}

The MIRI imaging data from Programmes~1269 and~4529 were simultaneously processed using the standard \emph{JWST} calibration pipeline (\textsc{jwst} v1.20.2) \citep{Bushouse2023}, with calibration context jwst\_1464.pmap. All individual files were processed through the initial stages of the pipeline, \texttt{calwebb\_detector1} and \texttt{calwebb\_image2}, following the default steps and pipeline configuration. The individually calibrated files were then combined into mosaics using the default processing steps in the \texttt{calwebb\_image3} stage of the pipeline. During the \texttt{calwebb\_image3} stage, the distortion-corrected exposures from different epochs were aligned and combined into a common astrometric frame; no additional image-registration or band-alignment steps were required. The resulting mosaics had excessively flagged outliers in some filters due to assumptions made by the \texttt{skymatch} step regarding the overlap area used to compute image statistics between pairs of exposures. This issue was particularly evident in the edge regions of individual F770W images.  Following consultation with the JWST Help Desk, the \texttt{skymatch} step was omitted from the processing. This substantially improved the quality of the resulting mosaics.

\begin{figure*}
    \centering
    \includegraphics[width=\textwidth]{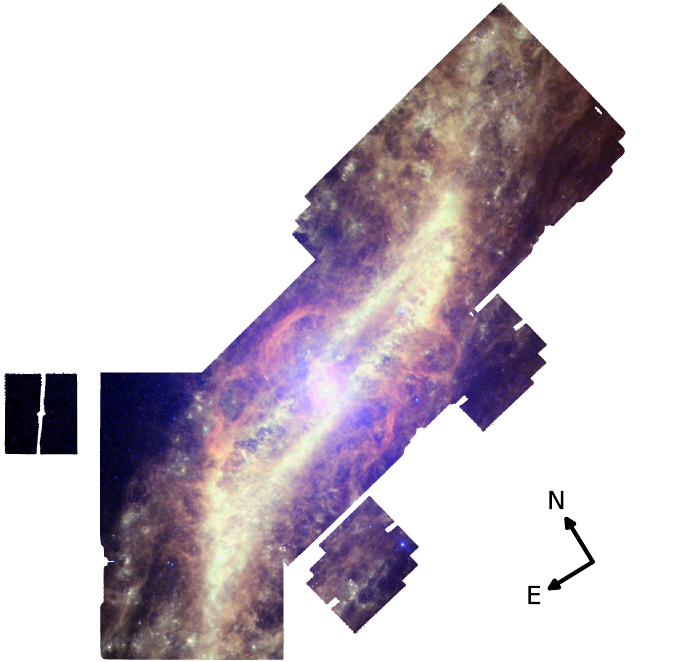}
    \caption{Three-colour \emph{JWST}/MIRI mosaic of the central region and inner disc of Centaurus~A. 
    The image combines F560W (blue), F770W (green), and F1130W (red), highlighting the prominent warped dust disc, filamentary structures, and the active nucleus. The nucleus appears as a bright compact source located between the prominent dust lanes of the warped disc. 
    The mosaic covers approximately $6.9~\mathrm{arcmin}^2$.} 
    \label{fig:miri_3color}
\end{figure*}

Figure~\ref{fig:miri_3color} shows a three-colour composite of the final MIRI mosaic of Centaurus~A, combining the F560W, F770W, and F1130W filters. The image highlights the complex morphology of the warped dust disc and associated filamentary structures that dominate the mid-infrared emission across the central regions of the galaxy (see Section~\ref{sec:structure}), as well as numerous sources distributed throughout the field of view. The active nucleus is visible at the centre of the mosaic as a bright, compact source located between the prominent dust lanes of the warped disc.

\subsection{Source Detection and Photometry}
\label{sec:Photometry}

Source detection and point-spread-function (PSF) photometry were performed using \starbug\ \citep{starbug2, Nally2024} version~\textit{v0.7.7}, a Python-based package optimised for JWST photometry in crowded fields and in spatially complex diffuse emission. \starbug\ has previously been applied to JWST observations of dust-rich environments in nearby galaxies ($D < 20$~Mpc), including studies of resolved stellar populations and deeply embedded low-mass sources \citep[e.g.,][]{Jones2023a, Habel2024, Lenkic2024, Zeidler2024, Yasui2026}.

Source detection was conducted on the Stage~3 mosaics in the MIRI F560W, F770W, and F1130W filters. Detections were performed independently in each band using the \starbug\ detection algorithm, which identifies statistically significant local maxima relative to a locally estimated background. This method is well suited to the Cen~A field, where strong and highly structured mid‑infrared emission from dust and polycyclic aromatic hydrocarbons (PAHs) produces a spatially variable background. The detection parameters adopted for each filter are summarised in Table~\ref{tab:starbug}. These were tuned separately for each band to account for background structure, enabling sensitivity to faint point sources while limiting contamination from structured peaks in the diffuse dust emission.

During the detection stage, aperture photometry was performed for all identified sources to provide initial flux estimates and to define the source lists used as positional priors for subsequent PSF fitting. The resulting catalogues were filtered using morphological criteria, including sharpness, roundness, and smoothness, to remove resolved sources. 

PSF photometry was performed on the individual Stage~2 calibrated exposures using the \starbug\ PSF-fitting routines. Performing the photometry on these exposures avoids correlated noise introduced during image mosaicking and preserves the native pixel sampling, improving positional accuracy. During the fitting procedure, small adjustments to source centroids relative to the mosaic-derived positions were permitted, while local background levels were simultaneously estimated and subtracted to mitigate the effects of diffuse emission. Photometry is reported in AB magnitudes using the standard JWST MIRI photometric zero-points provided by the pipeline calibration.

\begin{figure}
    \centering
    \includegraphics[width=\columnwidth]{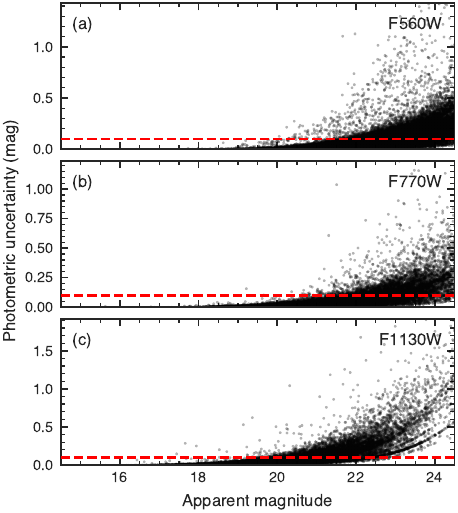}
    \caption{Photometric uncertainty as a function of AB magnitude for sources detected in the MIRI F560W, F770W, and F1130W filters. The Cen~A band-matched catalogue includes sources with uncertainties up to 0.4~mag. The red horizontal line indicates the more stringent 0.1~mag uncertainty threshold used in this work.}
    \label{fig:miri_mag_error}
\end{figure}

This process produced an independent catalogue for each dithered Stage~2 exposure in each filter. Within each band, these catalogues were spatially matched using the \starbug\ matching routines, adopting a maximum positional separation appropriate for the filter point-spread function. Sources detected in only a single exposure were rejected to suppress transient artefacts such as cosmic rays, and flux measurements from the remaining exposures were combined to produce a single per-band catalogue in AB magnitudes. The resulting photometric uncertainties as a function of magnitude for the MIRI F560W, F770W, and F1130W filters are shown in Figure~\ref{fig:miri_mag_error}.  

\begin{figure}
    \centering
    \includegraphics[width=\columnwidth]{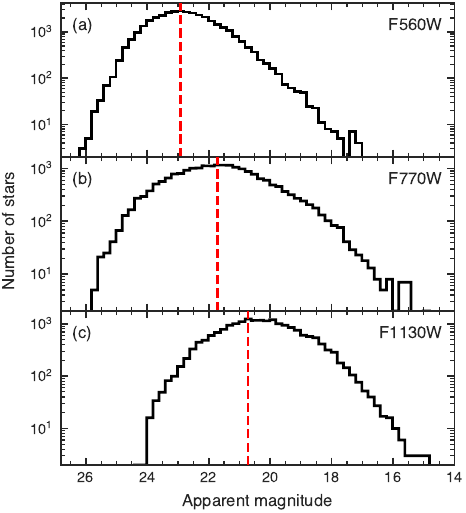}
    \caption{Luminosity functions for sources detected in the MIRI F560W, F770W, and F1130W filters. The turnover at faint magnitudes, marked by the red vertical dashed lines, indicates the estimated photometric completeness limit.}
    \label{fig:miri_luminosity_functions}
\end{figure}

The final band-matched photometric catalogue for Cen~A was constructed by matching sources across filters using their astrometric positions. Only sources with photometric uncertainties of $\leq 0.4$~mag in at least one band are included in the final catalogue.
The final catalogue contains a total of 58,445 sources, of which 4,745 are detected in all three MIRI bands. Of these, 2,558 have photometric uncertainties $\leq 0.1$~mag in all bands and are used for subsequent analysis.
This more stringent threshold ensures robust colour measurements for population separation, while the 0.4~mag limit preserves completeness in the catalogue.
Source coordinates were adopted from the shortest-wavelength filter in which each source was detected, reflecting the higher astrometric precision at shorter wavelengths. Table~\ref{tab:miri_catalogue_summary} summarises the number of
detections in each filter and the photometric completeness, derived from the
turnover of the luminosity functions in each band
(Figure~\ref{fig:miri_luminosity_functions}).

The contents of the final Cen~A band-matched catalogue are described in Table~\ref{tab:catalogue_columns}. The catalogue provides astrometric positions in the ICRS frame together with AB magnitudes and associated uncertainties for each MIRI filter. Magnitudes are not corrected for foreground or internal extinction. The full band-matched catalogue is released in electronic, machine-readable form with this paper.

\begin{table}
\centering
\caption{\starbug\ parameters used for aperture and point-spread function photometry in the MIRI observations of Cen~A. }
\label{tab:starbug}
\begin{tabular}{lccc}
\hline
Parameter     & F560W & F770W & F1130W \\
\hline
SIGSRC        & 3.5   & 3.2   & 3.0   \\
SIGSKY        & 1.5   & 1.5   & 1.5   \\
RICKER\_R     & 1.0   & 1.0   & 1.0   \\
SHARP\_LO     & 0.4   & 0.4   & 0.3   \\
SHARP\_HI     & 0.9   & 0.9   & 0.6   \\
ROUND\_LO/HI  & $\pm$1.0 & $\pm$1.0 & $+1.3/-1.4$ \\
\hline
APPHOT\_R     & 1.5   & 1.5   & 1.5   \\
SKY\_RIN      & 3.0   & 3.0   & 3.0   \\
SKY\_ROUT     & 4.5   & 4.5   & 4.5   \\
BOX\_SIZE     & 2     & 2     & 2     \\
CRIT\_SEP     & 6    & 6     & 6     \\
\hline
MATCH\_THRESH & 0.1   & 0.1   & 0.1   \\
NEXP\_THRESH  & All   & All   & All   \\
\hline
\end{tabular}
\end{table}

\begin{table}
\centering
\caption{Summary of the final band-matched PSF-photometry catalogues for the MIRI observations of Cen~A. Source counts are listed for each band; the final merged catalogue contains 58\,445 unique sources.}
\label{tab:miri_catalogue_summary}
\begin{tabular}{lcc}
\hline
Filter & Number of sources & Completeness limit (mag) \\
\hline
F560W  & 36754            & 22.91                     \\
F770W  & 21434            & 21.71                     \\
F1130W & 19838            & 20.71                     \\
\hline
\end{tabular}
\end{table}

\begin{table}
\centering
\caption{Description of columns in the final band-matched MIRI photometric catalogue for Cen~A. The full catalogue contains 58,445 sources with photometric uncertainties $\leq 0.4$ mag in at least one band; a more stringent $\leq 0.1$ mag cut is applied for analysis in this work.}
\label{tab:catalogue_columns}
\begin{tabular}{ll}
\hline
Column  & Description \\
\hline
CN      & Catalogue source identifier \\[2pt]
RA      & Right ascension (ICRS; degrees) \\
Dec     & Declination (ICRS; degrees) \\[2pt]
F560W   & AB magnitude in the MIRI F560W filter \\
eF560W  & Uncertainty on F560W magnitude \\[2pt]
F770W   & AB magnitude in the MIRI F770W filter \\
eF770W  & Uncertainty on F770W magnitude \\[2pt]
F1130W  & AB magnitude in the MIRI F1130W filter \\
eF1130W & Uncertainty on F1130W magnitude \\
\hline
\end{tabular}
\end{table}

\section{Results}
\label{Results}

\subsection{Resolved Stellar Populations}
\label{sec:stellar_pop}

The depth and angular resolution of the JWST/MIRI observations enable resolved mid-infrared point sources to be studied across the central regions of Cen~A. In this section, we use colour--magnitude (CMD) and colour--colour diagrams (CCD) constructed from the Cen~A point-source catalogue to identify distinct source populations and to isolate objects exhibiting infrared-excess emission. Unless otherwise stated, only sources with photometric uncertainties $\leq 0.1$~mag are included (see Section~\ref{sec:Photometry}). A total of 2,558 sources satisfy this criterion and are used for analysis. 

\subsubsection{MIRI Colour--Magnitude and Colour--Colour Diagrams}
\label{sec:CMDs}

\begin{figure}
    \centering
    \includegraphics[width=\columnwidth]{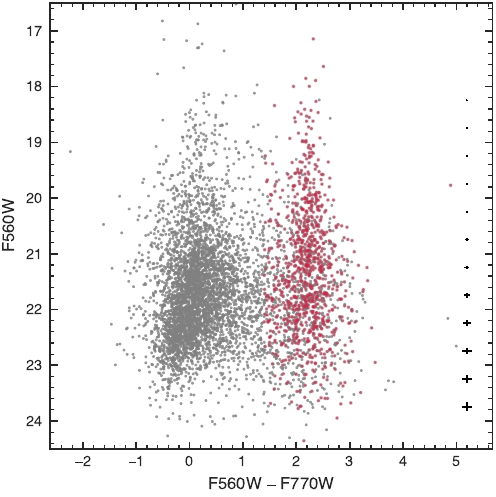}
    \caption{F560W versus F560W--F770W colour--magnitude diagram for Cen~A point sources. Grey points show all 2,558 sources; red points indicate the 928 objects with an infrared excess. Representative uncertainties as a function of magnitude are indicated on the right.}
    \label{fig:miri_57_cmd}
\end{figure}

\begin{figure}
    \centering
    \includegraphics[width=\columnwidth]{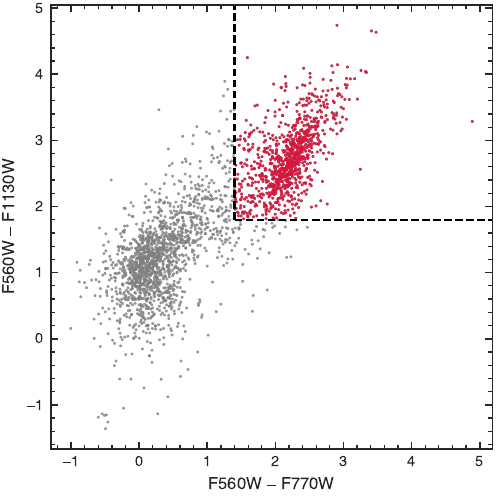}
    \caption{Colour--colour diagram of $F560W-F770W$ versus $F560W-F1130W$ for the Cen~A point-source sample. The dashed lines mark the adopted colour cuts used to select the red dust-enshrouded population. Symbols are colour-coded as in Fig.~\ref{fig:miri_57_cmd}}
    \label{fig:miri_ccd}
\end{figure}

Colour--magnitude and colour--colour diagrams constructed from the high-quality sample of 2,558 sources are shown in Figs~\ref{fig:miri_57_cmd} and \ref{fig:miri_ccd}.  The CMD plots F560W magnitude as a function of F560W--F770W 
colour, while the CCD shows F560W--F770W versus F560W--F1130W.

The F560W vs.~F560W--F770W CMD in Fig.~\ref{fig:miri_57_cmd} exhibits a clear bimodality. This indicates that two physically distinct populations are present, rather than a continuous distribution shaped solely by extinction.
The CMD is most densely populated near F560W--F770W $\simeq 0$, with these sources spanning the full range of F560W magnitudes. The relatively modest colour spread of this population is consistent with stellar photospheric emission \citep{Jones2017, Habel2024, Nayak2024b, Yasui2026}, with the scatter likely driven by spatially variable internal extinction within Cen~A. 
A second population is present at significantly redder colours, with a mean $\langle F560W{-}F770W \rangle = 2.20$~mag, while spanning a similar range in F560W magnitude. The red colours are indicative of strong dust emission, consistent with a substantial, physically distinct population of enshrouded sources in Cen~A. 
The separation is also apparent in the colour--colour plane (e.g., Fig.~\ref{fig:miri_ccd}). The resolved stellar population in Cen~A therefore includes both a population dominated by stellar photospheric emission and a significant red, dust-enshrouded population. The separation is less distinct in F770W--F1130W, where increased overlap likely reflects the contribution of PAH emission at longer wavelengths. These colours correspond to rising mid-infrared spectral slopes, indicative of warm dust emission rather than stellar photospheres. 

To select the red, dust-enshrouded sources identified in the Cen~A CMDs and CCDs, we require $F560W-F770W > 1.4$~mag and $F560W-F1130W > 1.8$~mag. Only sources with photometric uncertainties $\leq 0.1$~mag in all bands are included.
These cuts are chosen based on the minimum in source density separating the two populations in colour space. Similar separations between photospheric and dust-enshrouded sources are seen in JWST/MIRI studies of nearby star-forming regions \citep[e.g.,][]{Habel2024, Nayak2024b, Lenkic2024}, as well as in MIRI colour predictions for evolved stars and young stellar objects (YSOs)  based on radiative transfer models and empirical mid-IR spectra  \citep{Jones2017}. 
This selection yields a total of $N_{\rm IR}= 928$ red, dust-enshrouded sources, corresponding to 36.3\% of sources with $\sigma \leq 0.1$~mag in all three bands. These infrared-excess sources are highlighted as red points in Figs~\ref{fig:miri_57_cmd}  and \ref{fig:miri_ccd}.

\subsubsection{Spatial Distribution and Association with Dust Structures}
\label{sec:spatial_dist}

\begin{figure*}
    \centering
    \includegraphics[width=0.8\textwidth]{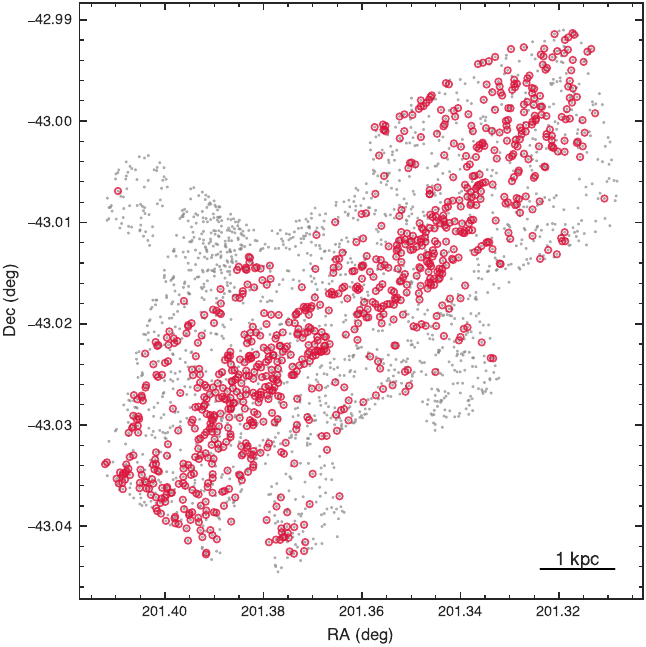}
    \caption{Spatial distribution of resolved sources in the central region of Cen~A. Grey points show all sources detected in all three MIRI bands with photometric uncertainties $<0.1$~mag, while red circles indicate the dust-enshrouded sources selected using the mid-infrared colour criteria.  A 1~kpc scale bar is shown.}   
    \label{fig:spatial_dist}
\end{figure*}

Having identified a distinct population of red, dust-enshrouded sources from their mid-infrared colours, we now examine their spatial distribution.
Figure~\ref{fig:spatial_dist} presents the spatial distribution of the stellar populations identified in the CMDs and CCDs across the central region of Cen~A, restricted to sources detected in all three MIRI bands with photometric uncertainties $<0.1$~mag. The red, dust-enshrouded sources closely trace the warped dust disc of Cen~A, in contrast to the more uniformly distributed population dominated by photospheric emission. This reveals distinct spatial distributions for the two populations and their differing associations with the galaxy’s large-scale dust structures.

\begin{figure*}
    \centering
    \includegraphics[width=0.95\textwidth, angle=0, trim=0mm 10mm 0mm 0mm, clip]{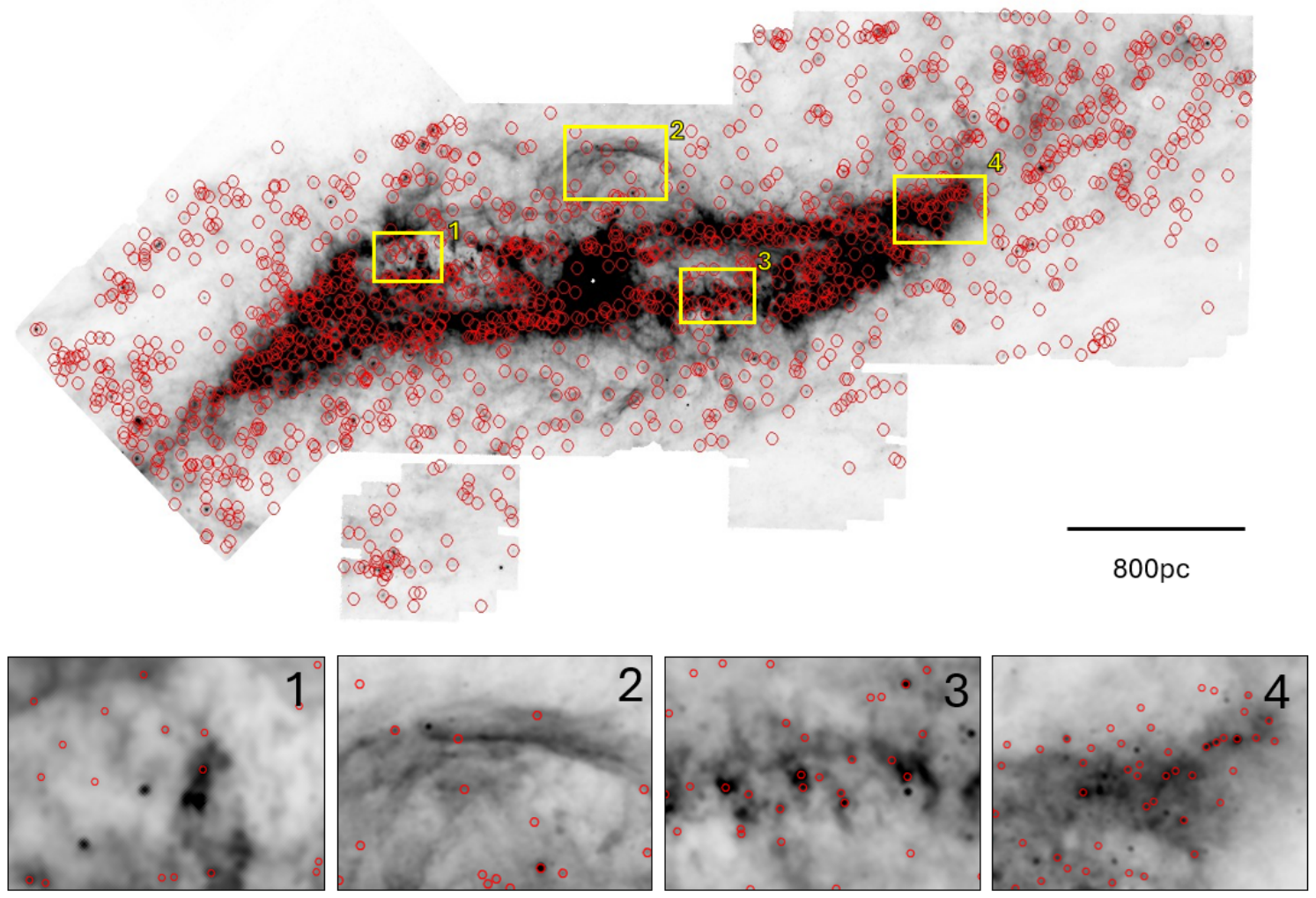}
    \caption{ Positions of the red, dust-enshrouded sources from Figure~\ref{fig:spatial_dist} overlaid on a wide-field MIRI F1130W image of the central region of Cen~A (top).
    Boxes in the top panel mark the locations of the zoomed-in regions, labelled 1–4. The lower panels sample a range of dust structures within the warped disc, with red circles indicating sources selected via the MIRI colour criteria. The red sources are strongly concentrated along the prominent dust lane and associated filamentary structures visible in the mid-infrared emission.
    }
    \label{fig:spatial_dist_image}
\end{figure*}

Figure~\ref{fig:spatial_dist_image} overlays the red, dust-enshrouded sources on the MIRI F1130W image, illustrating their close spatial association with the large-scale dust structures in the galaxy. The red sources are concentrated along the warped dust disc and its associated lanes and filaments visible in the mid-infrared, with relatively few located in regions largely devoid of diffuse dust emission.
The lower panels of Figure~\ref{fig:spatial_dist_image} show close-up views of regions surrounding representative red, dust-enshrouded sources (panels~1--4). The compact point sources are spatially coincident with narrow dust filaments and pillar-like features, often located at their tips or embedded within the brightest mid-infrared emission. Such morphologies are commonly observed in nearby star-forming regions, where embedded sources are found within dense, irradiated dust structures \citep{Reiter2022, Jones2023a, Andre2025, Zeidler2024}. 
The IR-excess sources (including candidate YSOs) broadly trace both the dusty filaments and the warped disc, consistent with star formation occurring in deeply embedded, dust- and molecular gas-rich environments. This spatial association is further supported by high-resolution CO observations, which reveal that the molecular gas in Cen~A is organised into filamentary structures throughout the dust lane \citep{Espada2019}, providing the dense environments required for embedded star formation.

A principal component analysis of the source positions (RA and Dec) demonstrates that the red, dust‑enshrouded sources form a strongly flattened distribution, with 94~per cent of the variance captured by the first principal component,  aligned with the major axis of the warped disc. In contrast, sources dominated by stellar photospheric emission are more broadly distributed, with only $\sim85$~per cent of the variance captured by the first component, reflecting a more spatially extended distribution. 
Consistent with this geometry, the red, dust-enshrouded sources are confined to a thinner structure, with a dispersion perpendicular to the disc of $\sigma_{\perp,\mathrm{red}} = 0.38$~kpc compared to $\sigma_{\perp,\mathrm{grey}} = 0.63$~kpc for the stellar photospheric population. This difference supports a thin, dust-embedded disc component superposed on a more vertically extended stellar distribution.

\subsection{Spectral Properties of the Red Source Population}
\label{sec:discussion_sed}

\begin{figure}
    \centering
    \includegraphics[width=\columnwidth]{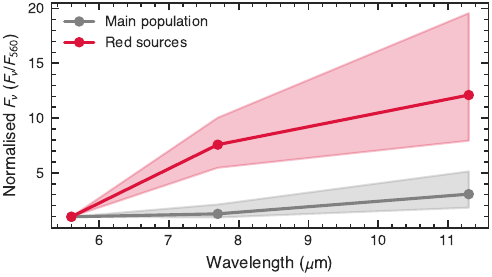}
    \caption{Median normalised mid-infrared SEDs of the colour-selected red sources (red) and the main point-source population (grey) in Cen~A. Fluxes are normalised at F560W (5.6~$\mu$m); shaded regions show the 16--84 per cent ranges. The red population exhibits a systematically steeper mid-infrared continuum consistent with emission dominated by warm dust.}
    \label{fig:sed_comparison}
\end{figure}

Figure~\ref{fig:sed_comparison} compares the median normalised mid-infrared spectral energy distributions (SEDs) of the colour-selected red sources and the main point-source population. The red sources exhibit systematically rising continua across the MIRI bands, in contrast to the comparatively flatter SEDs of the main population. The red sources have a median spectral slope of $\alpha = 3.55$ (16--84 per cent range: 2.95--4.24), compared to $\alpha = 1.60$ (0.86--2.33) for the main population, where $\alpha$ is defined between 5.6 and 11.3~$\mu$m. The weak rise observed in the median SED of the main population is likely driven by the increasing MIRI point-spread function with wavelength, which leads to residual inclusion of diffuse background emission. This includes low-level PAH contribution \citep{Leroy2023, Rigopoulou2024, Pantoni2026}, particularly in the F1130W band encompassing the 11.3~$\mu$m PAH feature, potentially enhancing the flux and thereby steepening the inferred slopes.

Extinction alone is unlikely to account for these steep spectral slopes. The mid-infrared extinction curve is relatively flat, with $A_{5.6\mu{\rm m}} \sim 0.05\,A_V$ and $A_{11.3\mu{\rm m}} \sim 0.02\,A_V$, implying $\Delta A \sim 0.03\,A_V$ across the MIRI bands. For typical values in Cen~A ($A_V \sim 7$--15), this corresponds to only $\sim0.2$--0.5~mag of reddening, insufficient to produce the strongly rising continua observed. The large positive slopes therefore require substantial intrinsic emission from warm circumstellar dust associated with embedded sources. In contrast, the more modest slopes of the main population ($\alpha \sim 1$--2) are consistent with predominantly photospheric emission, with the slight positive slopes arising from extinction and residual diffuse background emission.

\subsection{Dusty shell morphology and PAH properties}
\label{sec:structure}

The warped, nearly edge-on disc extends across the galactic nucleus along the east–west direction \citep{Quillen2006, Quillen2008}. Its distinct parallelogram morphology arises from line-of-sight projections of folds within a geometrically thin structure. Rich in molecular gas \citep{Quillen1992}, this region serves as the principal site of ongoing star formation within the galaxy \citep{Quillen1992, Mirabel1999, Leeuw2002, Quillen2006}.

The three-colour \emph{JWST} MIRI image (Fig.~\ref{fig:miri_3color}) reveals a system of extended mid-infrared structures, most prominently traced by the F1130W filter. These structures are also clearly visible in the F770W and even in the F560W images, although the emission becomes progressively fainter towards shorter wavelengths. They appear as a series of loop-like or bubble-like features located above and below the well-known parallelogram associated with the warped disc. Earlier \textit{Spitzer} observations with IRAC and IRS identified a related large-scale structure, referred to as an ``oval dusty shell'' \citep{Quillen2006, Quillen2008}, which was particularly prominent in the 11.3~$\mu$m PAH emission feature. The improved angular resolution of MIRI now resolves this structure into multiple distinct loops, indicating a significantly more complex morphology than previously recognised.

The dusty shells appear clearly distinct from the inner warped disc both morphologically and spectrally. In particular, previous work \citep{Quillen2008} has shown that the shell exhibits relatively enhanced 11.3~$\mu$m PAH emission compared to the 7.7~$\mu$m feature, implying elevated 11.3~$\mu$m to 7.7~$\mu$m PAH ratios relative to those measured in the star-forming regions of the parallelogram. This behaviour is suggestive of a PAH population that is more neutral, similar to that observed in outflow environments of nearby active galaxies \citep[e.g.][]{GarciaBernete2024_GATOS5}, as well as in the star-formation-driven wind of the nearby starburst galaxy M82 \citep{Cronin2026, Villanueva2025}, where \emph{JWST} observations show that PAH emission traces cold molecular gas within the outflow.

\begin{figure*}
    \centering
    \includegraphics[width=1.10\textwidth, trim=0 4.5cm 0 4.5cm, clip]{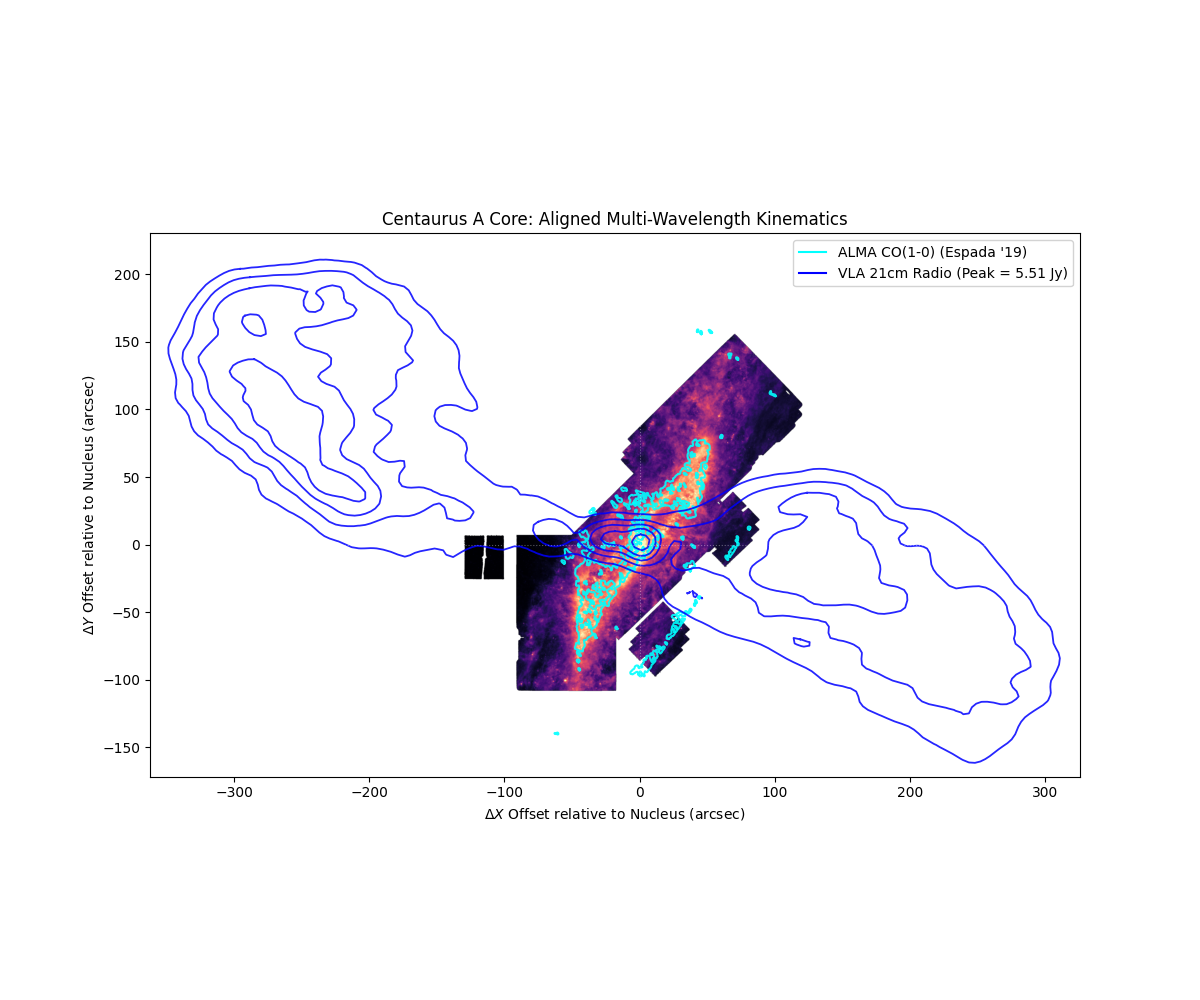}
    \caption{
    Comparison of the \emph{JWST}/MIRI F1130W emission with tracers of cold molecular gas and radio continuum in Cen~A. The \emph{JWST}/MIRI F1130W image is overlaid with contours of CO(1--0) emission \citep[cyan contours, from][with a beam of 1"$\times$1.5"]{Espada2019}, tracing the cold molecular gas, and VLA 21~cm radio continuum (blue contours) tracing large-scale radio emission \citep{Hardcastle2003}. The mid-infrared loop and filamentary structures closely follow the warped dust disc and molecular gas distribution, while showing little correspondence with the extended radio emission.
    }
    \label{fig:multiwavelength_overlay}
\end{figure*}

\begin{figure*}
    \centering
    \includegraphics[width=0.95\textwidth, trim=0 4.5cm 0 4.5cm, clip]{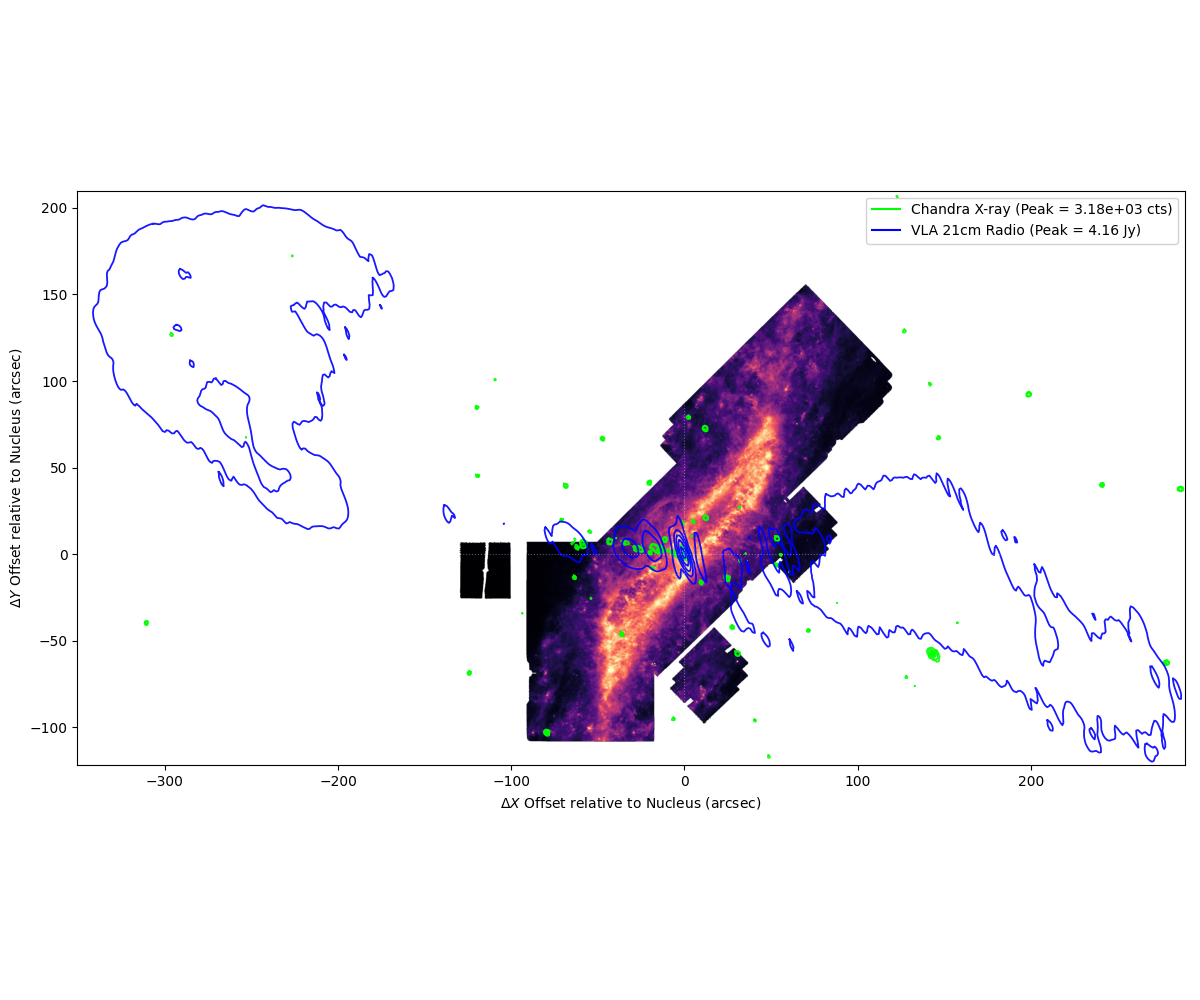}    
    \caption{
    Comparison of the \emph{JWST}/MIRI F1130W emission with radio and X-ray tracers of the AGN jet in Cen~A. The F1130W image is overlaid with VLA 21~cm radio continuum  \citep[blue contours;][]{Hardcastle2003} and \textit{Chandra} X-ray emission \citep[green contours;][]{Hardcastle2007_chandra}. The radio and X-ray emission trace the collimated AGN jet, which is oriented approximately perpendicular to the warped dust disc. The mid-infrared structures show no clear spatial correspondence with the jet.
    }
    \label{fig:jet_overlay}
\end{figure*}

The origin of these structures remains uncertain. Their morphology is reminiscent of large-scale bubble or shell features, analogous to the Galactic Centre hyper bubble. One possibility is that they are linked to energetic processes associated with the active nucleus, such as past episodes of jet activity or feedback.
This is illustrated in Fig.~\ref{fig:multiwavelength_overlay}, which compares the mid-infrared emission with tracers of the cold molecular gas (CO(1--0); \citealt{Espada2019}) and the large-scale radio jet \citep{Hardcastle2003}. The radio emission is highly collimated and does not align with the brightest mid-infrared loops, suggesting that the relationship between the dusty structures and the radio lobes is not clearly established. A similar lack of correspondence is seen in the distribution of ionised gas tracers such as [O~IV] \citep{Quillen2008}.
A comparison including X-ray emission tracing the hot gas associated with the jet is shown in Fig.~\ref{fig:jet_overlay}, reinforcing the lack of spatial correspondence between the jet and the mid-infrared structures.

The dusty shells are also known to be rich in molecular gas based on earlier infrared spectroscopy. Nevertheless, there is little evidence for significant ongoing star formation within these structures, as indicated by the apparent lack of embedded young sources in the images presented here. This combination of abundant dust and molecular gas with low star formation efficiency may indicate that the gas is dynamically heated and stabilised against collapse. {\em JWST}/MRS observations show that the molecular gas within the inner $\sim$100 pc is subject to strong turbulent dissipation and shock heating, maintaining temperatures above $\sim$100 K \citep{Evangelista2026}.  While these measurements probe much smaller scales than the shells, such conditions can inhibit gravitational collapse, thereby suppressing star formation despite the presence of substantial gas reservoirs.

\section{Discussion}
\label{Disc}

\subsection{Nature of the dust-enshrouded population}
The JWST/MIRI observations presented here reveal a distinct population of mid-infrared point sources with strong colour excesses that trace the warped dust disc of Cen~A. While their photometric properties clearly indicate emission dominated by warm dust, their physical nature cannot be uniquely determined from colours alone; however, spatial and demographic constraints strongly limit the range of viable interpretations.

In mid-IR colour space, embedded YSOs and evolved stars undergoing intense mass loss ($\dot{M} \gtrsim 10^{-5}\,\Msun\,\mathrm{yr}^{-1}$) can overlap. 
YSOs span a broad range of colours driven by envelope opacity, geometry, and evolutionary stage, while evolved stars follow more structured sequences governed primarily by dust chemistry and mass-loss rate \citep{Jones2017, Robitaille2006}.
The reddest colours are typically associated with heavily embedded, envelope-dominated YSOs, corresponding to early evolutionary stages (Stage~I; ages $\sim10^{5}$–$10^{6}$~yr), but can also be reached by the most extreme AGB stars during the short-lived superwind phase. 

The two populations seen in our MIRI colour--magnitude diagrams are also observed in JWST/MIRI studies of Local Group star-forming regions, where CMDs show a narrow sequence of evolved sources and a separate red population associated with embedded YSOs \citep{Lenkic2024, Habel2024, Nayak2024b}. In contrast, JWST studies of more quiescent, galaxy-wide stellar populations \citep[e.g.,][]{Nally2024} show a weaker separation and a less well-populated red component. The prominent red population in our MIRI CMDs is consistent with a population dominated by embedded YSO candidates and exhibits colours expected for envelope-dominated (Stage~I) YSOs, which typically have ages of $\sim10^{5}$--$10^{6}$~yr. 

These properties are consistent with the broader star-forming environment in Cen~A, in which the warped disc is rich in molecular gas \citep{Quillen1992, Quillen2008} and constitutes the primary site of ongoing star formation. Optical photometry has revealed increasingly blue colours towards the dust lane, a wide colour spread at its edges, and predominantly red colours within the dust lane itself \citep{vandenBergh1976}. This behaviour is attributed to a young stellar population emerging from dust, and is supported by the numerous H\,{\sc ii} regions in the central dust band \citep{Dufour1979,Hodge1983} and the very blue stellar populations at its northwestern and southeastern edges \citep{vandenBergh1976}. The strong spatial correlation between the red sources and filamentary dust structures is consistent with embedded star formation within dense, dusty environments.

Cen~A hosts a substantial population of evolved stars, including intermediate-age ($\sim2$--$4$~Gyr) asymptotic giant branch (AGB) stars \citep{Rejkuba2001,Rejkuba2022,Aghdam2024}. 
AGB stars trace older stellar populations ($\gtrsim30$~Myr), while red supergiants (RSGs) trace younger post-main-sequence populations ($\sim10$--$30$~Myr) and can also exhibit infrared excess.
Both phases experience strong mass loss, with rates up to $\sim10^{-5}\,M_\odot\,\mathrm{yr}^{-1}$, returning as much as $\sim80\%$ of their mass to the ISM through dust-driven winds \citep{Habing2003}.

AGB stars in Cen~A are expected to follow the underlying stellar mass distribution and therefore exhibit a relatively smooth spatial profile \citep{Rejkuba2022}. They would therefore follow the broader stellar component rather than being tightly confined to high-column density features, as observed here. 
Extreme, dust-enshrouded evolved stars are rare, comprising only $\sim$4~per~cent of the evolved stellar population in the Large Magellanic Cloud ($\sim$1340 out of $\sim3\times10^4$ sources; \citealt{Riebel2012, Srinivasan2009, Boyer2012, Jones2017_spec}). 
The number of red sources identified ($N_{\rm IR}=928$) far exceeds the expected population of extreme AGB stars based on scaling from Local Group systems.
Moreover, far-infrared surveys show that sources occupying the reddest regions of \textit{Spitzer} colour--magnitude space are dominated by YSOs, with only a negligible contribution from dusty evolved stars \citep{Seale2014, Jones2015}.
While an AGB population could exhibit some degree of flattening, it is unlikely to reproduce the highly flattened distribution, strong confinement to the warped disc, and association with filamentary dust structures observed here. This morphology implies that any contribution from evolved stars must be minor.

PAH features are rarely observed in AGB stars \citep{Sloan2007}, and in principle the 11.3~$\mu$m emission could help distinguish them from YSOs. However, strong diffuse PAH emission across the Cen~A disc means that the F1130W flux cannot be unambiguously attributed to individual point sources.

High-resolution $K$-band observations of the warped disc reveal bright, red sources associated with a ring structure, tentatively interpreted as RSGs or compact stellar clusters \citep{Kainulainen2009}. 
RSGs occupy only the brightest end of the luminosity distribution, whereas the red population identified in this work spans the full luminosity range sampled by MIRI/F560W in Cen~A. RSGs, therefore, cannot account for the observed red population. 

In addition to AGB and RSG stars, Wolf--Rayet (WR) stars are identified in Cen~A via He\,{\sc ii} and C\,{\sc iv} emission features in the spectra of young clusters \citep{Minniti2004}, tracing very recent ($\sim1$--10~Myr) massive star formation. While some WR systems (primarily WC+O binaries) are capable of producing substantial amounts of carbon-rich dust \citep{Williams1987}, this occurs in only a small fraction of the population and over a short-lived phase ($\sim10^5$~yr; \citealt{Crowther2007}). WR stars are therefore too rare to account for the large number of red sources observed across the warped disc. 

Background galaxies are unlikely to contribute significantly to the red population, given the strong spatial confinement of these sources to the warped dust disc. In addition, the photometric catalogue excludes resolved sources based on morphological criteria, further reducing contamination from high-redshift galaxies.

The colour distribution, spatial confinement to dense dust structures, and large source counts indicate that the red population is dominated by embedded YSO candidates, with only a minor contribution from evolved stars.

\subsection{Star formation within the warped dust disc}
To place the population of red, dust-enshrouded sources in the broader context of Cen~A, we consider the properties of the warped disc.

These sources are co-spatial with the warped dust disc in Cen~A, which is widely interpreted as the product of a gas-rich merger, with the accreted gas settling into the disc on a timescale of $\sim10^{8}$~yr \citep{Struve2010,Peng2002}. This event supplied a substantial reservoir of cold gas ($\sim10^{8}$--$10^{9}\,M_\odot$), which now supports star formation at a modest rate ($\lesssim 1\,M_\odot\,\mathrm{yr}^{-1}$), with efficiencies comparable to those in normal disc galaxies \citep{Espada2009,Espada2019,Wang2020}. 
 
Following the merger, the accreted gas cooled and settled into a rotating, warped disc traced by H\,{\sc i}, CO, and dust over scales from $\sim2$~pc to $\sim6.5$~kpc \citep{Wang2020,Quillen2010}. Neutral hydrogen observations indicate that the inner $\sim5$~kpc disc is largely settled and in regular rotation, with an age of a few $\times10^8$~yr \citep{Struve2010}, while more extended H\,\textsc{i} and molecular structures trace material that has not yet fully relaxed. The majority of the cold gas resides in the central disc ($M_{\mathrm{HI}} \sim 4.9 \times 10^8\,M_\odot$; $M_{\mathrm{H}_2} \sim 10^9$--$1.6 \times 10^9\,M_\odot$), while smaller fractions ($M_{\mathrm{HI}} \sim 5 \times 10^7\,M_\odot$) are found in shell- and ring-like structures extending to $\sim15$~kpc \citep{Wang2020,Espada2019}. This distribution indicates that the system is still dynamically evolving and has not yet fully relaxed following the merger.

The gas-rich warped disc hosts ongoing star formation, as indicated by H\,{\sc ii} regions and Pa$\alpha$ emission within the dust lane, implying circumnuclear star formation at a rate of $\sim0.1$--$1\,M_\odot\,\mathrm{yr}^{-1}$ \citep{Marconi2000,Espada2009}. This is consistent with the presence of substantial molecular gas reservoirs ($\sim10^9\,M_\odot$) and near-solar metallicities, indicating a chemically enriched ISM capable of sustaining long-lived, low-efficiency star formation within the settled merger remnant \citep{Espada2019,Peng2004b}. 

The red, dust-enshrouded sources identified above are consistent with a population of embedded YSOs ($\lesssim1$~Myr) associated with ongoing star formation within the warped dust disc of Cen~A. 

\subsection{Merger-driven versus jet-triggered star formation}
The origin of the red, dust-enshrouded population can be interpreted in terms of star formation regulated by the merger-accreted gas reservoir within the warped disc, or localised triggering by interactions between the AGN-driven radio jet and the surrounding ISM.
In addition, radio-mode AGN feedback can also suppress star formation by heating and turbulently stirring the ISM, preventing gas from collapsing \citep[e.g.,][]{Nesvadba2010}. However, we find no clear evidence for such suppression on the scales probed here.

The spatial distribution of the red sources provides a stringent geometric constraint on these scenarios. The embedded population forms a highly flattened structure aligned with the warped disc, with a small perpendicular dispersion (Section~\ref{sec:spatial_dist}), indicative of a geometrically thin structure. In contrast, the radio jet is inclined by $\sim70$--80$^\circ$ to the warped disc 
\citep{Croston2009,Wykes2013,Neff2015}. A jet-driven origin would be expected to produce a spatial correlation between the embedded sources and the jet axis; however, no such alignment or enhancement in source surface density is observed within the MIRI field. 

Indirect effects of the jet on the ISM (e.g. through pressure or turbulence) cannot be excluded, but are unlikely to dominate the distribution of star formation traced here. 
This is consistent with resolved molecular gas studies of Cen~A, which show that star formation efficiency is lower in the circumnuclear disc than in the outer regions, likely due to shear, turbulence, and AGN-related processes \citep{Espada2019}, supporting a scenario in which star formation is regulated by the properties of the disc gas rather than by the AGN.

This contrasts with the outer halo of Cen~A, where young stellar populations  coincide with filamentary structures that are aligned with the radio jet and are interpreted as jet-induced \citep{Mould2000,Crockett2012,Santoro2015,Rejkuba2022,Joseph2022}.  These regions are confined to large galactocentric radii ($\gtrsim10$--30~kpc) and organised into extended filaments, representing a localised and relatively low-efficiency mode of star formation \citep{Rejkuba2022,Oosterloo2005}. 
This is consistent with AGN-triggered star formation being confined to discrete regions rather than forming coherent large-scale structures \citep[e.g.,][]{Crockett2012,Cresci2015,Shin2019,Perna2020,HermosaMunoz2024b}, in contrast to the smooth, disc-wide distribution of embedded sources observed here.
The absence of any analogous alignment or anisotropy in the central few kiloparsecs likely rules out jet-driven triggering as the dominant mechanism forming the embedded population traced here. 

Instead, the embedded sources trace the high-column density structures of the warped disc, closely following its filamentary morphology. This indicates that star formation is governed by the distribution of cold molecular gas deposited during the merger and subsequently settled into the disc, consistent with H\,\textsc{i} studies showing that the merger predates and is not directly linked to the current AGN activity \citep{Struve2010}. This interpretation is supported by evidence for enhanced star formation across Cen~A around $\sim800$~Myr ago, consistent with a merger event that supplied the gas reservoir from which the present disc formed \citep{Aghdam2024}. 

Although radio jets with powers comparable to Cen~A ($\sim10^{43}$\,erg\,s$^{-1}$; \citealt{Israel1998}) can drive shocks and redistribute gas \citep{Mukherjee2016,Mukherjee2018,Talbot2022}, jet--ISM interactions typically produce turbulent, inefficient, and spatially intermittent star formation rather than coherent large-scale structures \citep{Salome2017}.

At smaller scales, however, \emph{JWST}/MIRI MRS observations reveal clear signatures of jet--ISM interaction within the inner $\sim$100--200~pc of Cen~A. The ionised gas shows fast outflows and shock-excited emission perpendicular to the radio jet, consistent with a jet-inflated bubble interacting with the surrounding ISM \citep{Mukherjee2018,AlonsoHerrero2025}, while the warm molecular gas exhibits disturbed morphologies and localised excitation linked to shocks \citep{Evangelista2026}. Nevertheless, the coupling between the jet and the molecular disc appears limited, with no strong molecular outflow detected and only localised perturbations in the warm gas. These effects are therefore confined to the circumnuclear region and do not account for the large-scale distribution of embedded sources. 

\subsection{Caveats and limitations} 
Distinguishing between embedded YSOs and dust-producing evolved stars remains challenging, and some contamination from AGB stars is likely. The classification presented here is based on mid-infrared colours and spatial associations and lacks direct spectroscopic confirmation. Uncertainties in extinction, intrinsic luminosities, and selection effects associated with the adopted colour criteria may also influence the inferred properties of the red source population.

The strong confinement of the red, dust-enshrouded sources to the warped dust disc of Cen~A, rather than alignment with the AGN jet, supports an origin in young, embedded stellar objects formed within the disc. Given that the dust disc itself is widely interpreted as the product of a past merger, the observed star formation is likely associated with the accreted gas reservoir rather than being directly triggered by jet--ISM interactions. While our analysis demonstrates that the red sources form a distinct population within the warped dust disc, their intrinsic luminosities, extinction, and immediate environments cannot be constrained using mid-infrared imaging alone.
Observations at near-infrared wavelengths can provide constraints on the underlying stellar emission, while molecular-line observations in the submillimetre and spectroscopic follow-up will directly characterise the associated gas and assess the nature of the embedded star formation.

\section{Summary and Conclusions}
\label{Summ_Conc}

We present \emph{JWST}/MIRI F560W, F770W and F1130W imaging of the central regions of Centaurus~A, covering $\sim4\times2$ kpc, resolving its mid-infrared point-source population within the warped dust disc. The resulting band-matched catalogue contains 58,445 sources. This provides the first resolved view of stellar populations within the warped dust disc and the immediate vicinity of the AGN. A distinct population of 928 red sources with strong infrared excess is identified from colour–magnitude and colour–colour diagnostics. The imaging also reveals that the dusty shell previously identified with \textit{Spitzer} is composed of multiple loop-like structures associated with the warped disc, with emission strongest in F1130W.

The red sources trace the warped dust disc and its filamentary structures, and are consistent with a population dominated by embedded YSO candidates in dense, dust-rich environments. Their colours and spatial distribution indicate very recent star formation within the disc, on timescales of a few Myr. The lack of alignment with the radio jet indicates that ongoing star formation in Cen~A is primarily associated with the merger-accreted gas reservoir rather than being driven by jet--ISM interactions. 
These results demonstrate that Cen~A continues to sustain embedded star formation within its merger‑accreted disc, despite hosting an active nucleus.
Further constraints on the nature of these sources will require complementary near-infrared imaging and spectroscopic follow-up. 
The band-matched MIRI photometric catalogue is made publicly available.


\vspace{2ex}

\noindent {\it Facilities:} {\em JWST} (MIRI) - James Webb Space Telescope.

\noindent {\it Software:} {\sc photutils} \citep{photutils},  \starbug\ \citep{starbug2, Nally2024}, and {\sc topcat} \citep{Taylor2005}.

\section*{Acknowledgements}

This work is based on observations made with the NASA/ESA/CSA James Webb Space Telescope. The data were obtained from the Mikulski Archive for Space Telescopes at the Space Telescope Science Institute, which is operated by the Association of Universities for Research in Astronomy, Inc., under NASA contract NAS 5-03127 for JWST. These observations are associated with programs \#1269 and 4529. This work is based in part on observations made with the Spitzer Space Telescope, which was operated by the Jet Propulsion Laboratory, California Institute of Technology, under a contract with NASA.

O.C.J. acknowledges support from an STFC Webb fellowship.
AAH acknowledges support from grant PID2021-124665NB-I00  funded by MCIN/AEI/10.13039/501100011033 and by ERDF A way of making Europe.
KJ acknowledges the support from the Swedish National Space Agency.
MM acknowledges support through a NASA-JPL/JWST task plan 71-209636. MM also acknowledges that a portion of this research was carried out at the Jet Propulsion Laboratory, California Institute of Technology, under contract with the National Aeronautics and Space Administration (80NM0018D0004).
LP and MB acknowledge funding from the Belgian Science Policy Office (BELSPO) through the PRODEX project “JWST/MIRI Science exploitation” (C4000142239). 
LC acknowledges support from grant PID2021-127718NB-100 from the Spanish Ministry of Science and Innovation/State Agency of Research MCIN/AEI/10.13039/501100011033 and by “ERDF A way of making Europe”


\section*{Data availability}

The \emph{JWST} data underlying this article are publicly available from the
Mikulski Archive for Space Telescopes (MAST;
\url{https://archive.stsci.edu/}) under Programmes 1269 and 4529.

The band-matched \emph{JWST}/MIRI point-source catalogue presented in this
work is available as supplementary material accompanying the published
article. The catalogue contains astrometric positions and photometry in the
F560W, F770W, and F1130W filters for 58\,445 sources detected in the central
regions of Centaurus~A.



\bibliographystyle{mnras}

\begin{thebibliography}{}
\makeatletter
\relax
\def\mn@urlcharsother{\let\do\@makeother \do\$\do\&\do\#\do\^\do\_\do\%\do\~}
\def\mn@doi{\begingroup\mn@urlcharsother \@ifnextchar [ {\mn@doi@}
  {\mn@doi@[]}}
\def\mn@doi@[#1]#2{\def\@tempa{#1}\ifx\@tempa\@empty \href
  {http://dx.doi.org/#2} {doi:#2}\else \href {http://dx.doi.org/#2} {#1}\fi
  \endgroup}
\def\mn@eprint#1#2{\mn@eprint@#1:#2::\@nil}
\def\mn@eprint@arXiv#1{\href {http://arxiv.org/abs/#1} {{\tt arXiv:#1}}}
\def\mn@eprint@dblp#1{\href {http://dblp.uni-trier.de/rec/bibtex/#1.xml}
  {dblp:#1}}
\def\mn@eprint@#1:#2:#3:#4\@nil{\def\@tempa {#1}\def\@tempb {#2}\def\@tempc
  {#3}\ifx \@tempc \@empty \let \@tempc \@tempb \let \@tempb \@tempa \fi \ifx
  \@tempb \@empty \def\@tempb {arXiv}\fi \@ifundefined
  {mn@eprint@\@tempb}{\@tempb:\@tempc}{\expandafter \expandafter \csname
  mn@eprint@\@tempb\endcsname \expandafter{\@tempc}}}

\bibitem[\protect\citeauthoryear{{Aghdam} et~al.,}{{Aghdam}
  et~al.}{2024}]{Aghdam2024}
{Aghdam} S.~T.,  et~al., 2024, \mn@doi [\apj] {10.3847/1538-4357/ad57c0}, \href
  {https://ui.adsabs.harvard.edu/abs/2024ApJ...972...47A} {972, 47}

\bibitem[\protect\citeauthoryear{{Alonso Herrero} et~al.,}{{Alonso Herrero}
  et~al.}{2025}]{AlonsoHerrero2025}
{Alonso Herrero} A.,  et~al., 2025, \mn@doi [\aap]
  {10.1051/0004-6361/202554823}, \href
  {https://ui.adsabs.harvard.edu/abs/2025A&A...699A.334A} {699, A334}

\bibitem[\protect\citeauthoryear{{Andr{\'e}}, {Mattern}, {Arzoumanian},
  {Shimajiri}, {Zavagno}, {Abe}  \& {Russeil}}{{Andr{\'e}}
  et~al.}{2025}]{Andre2025}
{Andr{\'e}} P.,  {Mattern} M.,  {Arzoumanian} D.,  {Shimajiri} Y.,  {Zavagno}
  A.,  {Abe} D.,   {Russeil} D.,  2025, \mn@doi [\apjl]
  {10.3847/2041-8213/adc73d}, \href
  {https://ui.adsabs.harvard.edu/abs/2025ApJ...984L..59A} {984, L59}

\bibitem[\protect\citeauthoryear{{Boyer} et~al.,}{{Boyer}
  et~al.}{2012}]{Boyer2012}
{Boyer} M.~L.,  et~al., 2012, \mn@doi [\apj] {10.1088/0004-637X/748/1/40},
  \href {https://ui.adsabs.harvard.edu/abs/2012ApJ...748...40B} {748, 40}

\bibitem[\protect\citeauthoryear{Bradley et~al.,}{Bradley
  et~al.}{2024}]{photutils}
Bradley L.,  et~al., 2024, astropy/photutils: 2.0.2,
  \mn@doi{10.5281/zenodo.13989456}, \url
  {https://doi.org/10.5281/zenodo.13989456}

\bibitem[\protect\citeauthoryear{{Bushouse} et~al.,}{{Bushouse}
  et~al.}{2025}]{Bushouse2023}
{Bushouse} H.,  et~al., 2025, {JWST Calibration Pipeline},
  \mn@doi{10.5281/zenodo.6984365}

\bibitem[\protect\citeauthoryear{{Clarke}, {Burns}  \& {Norman}}{{Clarke}
  et~al.}{1992}]{Clarke1992}
{Clarke} D.~A.,  {Burns} J.~O.,   {Norman} M.~L.,  1992, \mn@doi [\apj]
  {10.1086/171663}, \href
  {https://ui.adsabs.harvard.edu/abs/1992ApJ...395..444C} {395, 444}

\bibitem[\protect\citeauthoryear{{Cresci} et~al.,}{{Cresci}
  et~al.}{2015}]{Cresci2015}
{Cresci} G.,  et~al., 2015, \mn@doi [\apj] {10.1088/0004-637X/799/1/82}, \href
  {https://ui.adsabs.harvard.edu/abs/2015ApJ...799...82C} {799, 82}

\bibitem[\protect\citeauthoryear{{Crockett} et~al.,}{{Crockett}
  et~al.}{2012}]{Crockett2012}
{Crockett} R.~M.,  et~al., 2012, \mn@doi [\mnras]
  {10.1111/j.1365-2966.2012.20418.x}, \href
  {https://ui.adsabs.harvard.edu/abs/2012MNRAS.421.1603C} {421, 1603}

\bibitem[\protect\citeauthoryear{{Cronin} et~al.,}{{Cronin}
  et~al.}{2026}]{Cronin2026}
{Cronin} S.~A.,  et~al., 2026, \mn@doi [\apj] {10.3847/1538-4357/ae5f66}, \href
  {https://ui.adsabs.harvard.edu/abs/2026ApJ..1002..217C} {1002, 217}

\bibitem[\protect\citeauthoryear{{Croston} et~al.,}{{Croston}
  et~al.}{2009}]{Croston2009}
{Croston} J.~H.,  et~al., 2009, \mn@doi [\mnras]
  {10.1111/j.1365-2966.2009.14715.x}, \href
  {https://ui.adsabs.harvard.edu/abs/2009MNRAS.395.1999C} {395, 1999}

\bibitem[\protect\citeauthoryear{{Crowther}}{{Crowther}}{2007}]{Crowther2007}
{Crowther} P.~A.,  2007, \mn@doi [\araa]
  {10.1146/annurev.astro.45.051806.110615}, \href
  {https://ui.adsabs.harvard.edu/abs/2007ARA&A..45..177C} {45, 177}

\bibitem[\protect\citeauthoryear{{Dicken} et~al.,}{{Dicken}
  et~al.}{2024}]{Dicken2024}
{Dicken} D.,  et~al., 2024, \mn@doi [\aap] {10.1051/0004-6361/202449451}, \href
  {https://ui.adsabs.harvard.edu/abs/2024A&A...689A...5D} {689, A5}

\bibitem[\protect\citeauthoryear{{Dufour}, {van den Bergh}, {Harvel},
  {Martins}, {Schiffer}, {Talbot}, {Talent}  \& {Wells}}{{Dufour}
  et~al.}{1979}]{Dufour1979}
{Dufour} R.~J.,  {van den Bergh} S.,  {Harvel} C.~A.,  {Martins} D.~H.,
  {Schiffer} III F.~H.,  {Talbot} Jr. R.~J.,  {Talent} D.~L.,   {Wells} D.~C.,
  1979, \mn@doi [\aj] {10.1086/112421}, \href
  {https://ui.adsabs.harvard.edu/abs/1979AJ.....84..284D} {84, 284}

\bibitem[\protect\citeauthoryear{{Espada} et~al.,}{{Espada}
  et~al.}{2009}]{Espada2009}
{Espada} D.,  et~al., 2009, \mn@doi [\apj] {10.1088/0004-637X/695/1/116}, \href
  {https://ui.adsabs.harvard.edu/abs/2009ApJ...695..116E} {695, 116}

\bibitem[\protect\citeauthoryear{{Espada} et~al.,}{{Espada}
  et~al.}{2017}]{Espada2017}
{Espada} D.,  et~al., 2017, \mn@doi [\apj] {10.3847/1538-4357/aa78a9}, \href
  {https://ui.adsabs.harvard.edu/abs/2017ApJ...843..136E} {843, 136}

\bibitem[\protect\citeauthoryear{{Espada} et~al.,}{{Espada}
  et~al.}{2019}]{Espada2019}
{Espada} D.,  et~al., 2019, \mn@doi [\apj] {10.3847/1538-4357/ab262d}, \href
  {https://ui.adsabs.harvard.edu/abs/2019ApJ...887...88E} {887, 88}

\bibitem[\protect\citeauthoryear{{Evangelista} et~al.,}{{Evangelista}
  et~al.}{2026}]{Evangelista2026}
{Evangelista} L.,  et~al., 2026, \mn@doi [arXiv e-prints]
  {10.48550/arXiv.2605.22497}, \href
  {https://ui.adsabs.harvard.edu/abs/2026arXiv260522497E} {p. arXiv:2605.22497}

\bibitem[\protect\citeauthoryear{{Garc{\'\i}a-Bernete}
  et~al.,}{{Garc{\'\i}a-Bernete} et~al.}{2024}]{GarciaBernete2024_GATOS5}
{Garc{\'\i}a-Bernete} I.,  et~al., 2024, \mn@doi [\aap]
  {10.1051/0004-6361/202450086}, \href
  {https://ui.adsabs.harvard.edu/abs/2024A&A...691A.162G} {691, A162}

\bibitem[\protect\citeauthoryear{{Gardner} et~al.,}{{Gardner}
  et~al.}{2023}]{Gardner2023}
{Gardner} J.~P.,  et~al., 2023, \mn@doi [\pasp] {10.1088/1538-3873/acd1b5},
  \href {https://ui.adsabs.harvard.edu/abs/2023PASP..135f8001G} {135, 068001}

\bibitem[\protect\citeauthoryear{{Graham}}{{Graham}}{1979}]{Graham1979}
{Graham} J.~A.,  1979, \mn@doi [\apj] {10.1086/157265}, \href
  {https://ui.adsabs.harvard.edu/abs/1979ApJ...232...60G} {232, 60}

\bibitem[\protect\citeauthoryear{{Habel} et~al.,}{{Habel}
  et~al.}{2024}]{Habel2024}
{Habel} N.,  et~al., 2024, \mn@doi [\apj] {10.3847/1538-4357/ad5343}, \href
  {https://ui.adsabs.harvard.edu/abs/2024ApJ...971..108H} {971, 108}

\bibitem[\protect\citeauthoryear{{Habing} \& {Olofsson}}{{Habing} \&
  {Olofsson}}{2003}]{Habing2003}
{Habing} H.~J.,  {Olofsson} H.,  eds, 2003, {Asymptotic giant branch stars}

\bibitem[\protect\citeauthoryear{{Hardcastle}, {Worrall}, {Kraft}, {Forman},
  {Jones}  \& {Murray}}{{Hardcastle} et~al.}{2003}]{Hardcastle2003}
{Hardcastle} M.~J.,  {Worrall} D.~M.,  {Kraft} R.~P.,  {Forman} W.~R.,  {Jones}
  C.,   {Murray} S.~S.,  2003, \mn@doi [\apj] {10.1086/376519}, \href
  {https://ui.adsabs.harvard.edu/abs/2003ApJ...593..169H} {593, 169}

\bibitem[\protect\citeauthoryear{{Hardcastle} et~al.,}{{Hardcastle}
  et~al.}{2007}]{Hardcastle2007_chandra}
{Hardcastle} M.~J.,  et~al., 2007, \mn@doi [\apjl] {10.1086/524197}, \href
  {https://ui.adsabs.harvard.edu/abs/2007ApJ...670L..81H} {670, L81}

\bibitem[\protect\citeauthoryear{{Harris}}{{Harris}}{2010}]{Harris2010}
{Harris} G. L.~H.,  2010, \mn@doi [\pasa] {10.1071/AS09063}, \href
  {https://ui.adsabs.harvard.edu/abs/2010PASA...27..475H} {27, 475}

\bibitem[\protect\citeauthoryear{{Harris} \& {Harris}}{{Harris} \&
  {Harris}}{2002}]{Harris2002}
{Harris} W.~E.,  {Harris} G. L.~H.,  2002, \mn@doi [\aj] {10.1086/340466},
  \href {https://ui.adsabs.harvard.edu/abs/2002AJ....123.3108H} {123, 3108}

\bibitem[\protect\citeauthoryear{{Hermosa Mu{\~n}oz} et~al.,}{{Hermosa
  Mu{\~n}oz} et~al.}{2024}]{HermosaMunoz2024b}
{Hermosa Mu{\~n}oz} L.,  et~al., 2024, \mn@doi [\aap]
  {10.1051/0004-6361/202450262}, \href
  {https://ui.adsabs.harvard.edu/abs/2024A&A...690A.350H} {690, A350}

\bibitem[\protect\citeauthoryear{{Hodge} \& {Kennicutt}}{{Hodge} \&
  {Kennicutt}}{1983}]{Hodge1983}
{Hodge} P.~W.,  {Kennicutt} Jr. R.~C.,  1983, \mn@doi [\aj] {10.1086/113318},
  \href {https://ui.adsabs.harvard.edu/abs/1983AJ.....88..296H} {88, 296}

\bibitem[\protect\citeauthoryear{{Israel}}{{Israel}}{1998}]{Israel1998}
{Israel} F.~P.,  1998, \mn@doi [\aapr] {10.1007/s001590050011}, \href
  {https://ui.adsabs.harvard.edu/abs/1998A&ARv...8..237I} {8, 237}

\bibitem[\protect\citeauthoryear{{Jones}, {Meixner}, {Sargent}, {Boyer},
  {Sewi{\l}o}, {Hony}  \& {Roman-Duval}}{{Jones} et~al.}{2015}]{Jones2015}
{Jones} O.~C.,  {Meixner} M.,  {Sargent} B.~A.,  {Boyer} M.~L.,  {Sewi{\l}o}
  M.,  {Hony} S.,   {Roman-Duval} J.,  2015, \mn@doi [\apj]
  {10.1088/0004-637X/811/2/145}, \href
  {https://ui.adsabs.harvard.edu/abs/2015ApJ...811..145J} {811, 145}

\bibitem[\protect\citeauthoryear{{Jones} et~al.,}{{Jones}
  et~al.}{2017a}]{Jones2017_spec}
{Jones} O.~C.,  et~al., 2017a, \mn@doi [\mnras] {10.1093/mnras/stx1101}, \href
  {https://ui.adsabs.harvard.edu/abs/2017MNRAS.470.3250J} {470, 3250}

\bibitem[\protect\citeauthoryear{{Jones}, {Meixner}, {Justtanont}  \&
  {Glasse}}{{Jones} et~al.}{2017b}]{Jones2017}
{Jones} O.~C.,  {Meixner} M.,  {Justtanont} K.,   {Glasse} A.,  2017b, \mn@doi
  [\apj] {10.3847/1538-4357/aa6bf6}, \href
  {https://ui.adsabs.harvard.edu/abs/2017ApJ...841...15J} {841, 15}

\bibitem[\protect\citeauthoryear{{Jones} et~al.,}{{Jones}
  et~al.}{2023}]{Jones2023a}
{Jones} O.~C.,  et~al., 2023, \mn@doi [Nature Astronomy]
  {10.1038/s41550-023-01945-7}, \href
  {https://ui.adsabs.harvard.edu/abs/2023NatAs...7..694J} {7, 694}

\bibitem[\protect\citeauthoryear{{Joseph}, {Sreekumar}, {Stalin}, {Paul},
  {Mondal}, {George}  \& {Mathew}}{{Joseph} et~al.}{2022}]{Joseph2022}
{Joseph} P.,  {Sreekumar} P.,  {Stalin} C.~S.,  {Paul} K.~T.,  {Mondal} C.,
  {George} K.,   {Mathew} B.,  2022, \mn@doi [\mnras] {10.1093/mnras/stac2388},
  \href {https://ui.adsabs.harvard.edu/abs/2022MNRAS.516.2300J} {516, 2300}

\bibitem[\protect\citeauthoryear{{Kainulainen} et~al.,}{{Kainulainen}
  et~al.}{2009}]{Kainulainen2009}
{Kainulainen} J.~T.,  et~al., 2009, \mn@doi [\aap]
  {10.1051/0004-6361/200912624}, \href
  {https://ui.adsabs.harvard.edu/abs/2009A&A...502L...5K} {502, L5}

\bibitem[\protect\citeauthoryear{{Keel}, {Banfield}, {Medling}  \&
  {Neff}}{{Keel} et~al.}{2019}]{Keel2019}
{Keel} W.~C.,  {Banfield} J.~K.,  {Medling} A.~M.,   {Neff} S.~G.,  2019,
  \mn@doi [\aj] {10.3847/1538-3881/aaf809}, \href
  {https://ui.adsabs.harvard.edu/abs/2019AJ....157...66K} {157, 66}

\bibitem[\protect\citeauthoryear{{Leeuw}, {Hawarden}, {Matthews}, {Robson}  \&
  {Eckart}}{{Leeuw} et~al.}{2002}]{Leeuw2002}
{Leeuw} L.~L.,  {Hawarden} T.~G.,  {Matthews} H.~E.,  {Robson} E.~I.,
  {Eckart} A.,  2002, \mn@doi [\apj] {10.1086/324494}, \href
  {https://ui.adsabs.harvard.edu/abs/2002ApJ...565..131L} {565, 131}

\bibitem[\protect\citeauthoryear{{Lenki{\'c}} et~al.,}{{Lenki{\'c}}
  et~al.}{2024}]{Lenkic2024}
{Lenki{\'c}} L.,  et~al., 2024, \mn@doi [\apj] {10.3847/1538-4357/ad3f90},
  \href {https://ui.adsabs.harvard.edu/abs/2024ApJ...967..110L} {967, 110}

\bibitem[\protect\citeauthoryear{{Leroy} et~al.,}{{Leroy}
  et~al.}{2023}]{Leroy2023}
{Leroy} A.~K.,  et~al., 2023, \mn@doi [\apjl] {10.3847/2041-8213/acaf85}, \href
  {https://ui.adsabs.harvard.edu/abs/2023ApJ...944L...9L} {944, L9}

\bibitem[\protect\citeauthoryear{{Marconi}, {Schreier}, {Koekemoer}, {Capetti},
  {Axon}, {Macchetto}  \& {Caon}}{{Marconi} et~al.}{2000}]{Marconi2000}
{Marconi} A.,  {Schreier} E.~J.,  {Koekemoer} A.,  {Capetti} A.,  {Axon} D.,
  {Macchetto} D.,   {Caon} N.,  2000, \mn@doi [\apj] {10.1086/308168}, \href
  {https://ui.adsabs.harvard.edu/abs/2000ApJ...528..276M} {528, 276}

\bibitem[\protect\citeauthoryear{{Minniti}, {Rejkuba}, {Funes}  \&
  {Kennicutt}}{{Minniti} et~al.}{2004}]{Minniti2004}
{Minniti} D.,  {Rejkuba} M.,  {Funes} J.~G.,   {Kennicutt} Jr. R.~C.,  2004,
  \mn@doi [\apj] {10.1086/422546}, \href
  {https://ui.adsabs.harvard.edu/abs/2004ApJ...612..215M} {612, 215}

\bibitem[\protect\citeauthoryear{{Mirabel} et~al.,}{{Mirabel}
  et~al.}{1999}]{Mirabel1999}
{Mirabel} I.~F.,  et~al., 1999, \mn@doi [\aap]
  {10.48550/arXiv.astro-ph/9810419}, \href
  {https://ui.adsabs.harvard.edu/abs/1999A&A...341..667M} {341, 667}

\bibitem[\protect\citeauthoryear{{Mould} et~al.,}{{Mould}
  et~al.}{2000}]{Mould2000}
{Mould} J.~R.,  et~al., 2000, \mn@doi [\apj] {10.1086/308927}, \href
  {https://ui.adsabs.harvard.edu/abs/2000ApJ...536..266M} {536, 266}

\bibitem[\protect\citeauthoryear{{Mukherjee}, {Bicknell}, {Sutherland}  \&
  {Wagner}}{{Mukherjee} et~al.}{2016}]{Mukherjee2016}
{Mukherjee} D.,  {Bicknell} G.~V.,  {Sutherland} R.,   {Wagner} A.,  2016,
  \mn@doi [\mnras] {10.1093/mnras/stw1368}, \href
  {https://ui.adsabs.harvard.edu/abs/2016MNRAS.461..967M} {461, 967}

\bibitem[\protect\citeauthoryear{{Mukherjee}, {Bicknell}, {Wagner},
  {Sutherland}  \& {Silk}}{{Mukherjee} et~al.}{2018}]{Mukherjee2018}
{Mukherjee} D.,  {Bicknell} G.~V.,  {Wagner} A.~Y.,  {Sutherland} R.~S.,
  {Silk} J.,  2018, \mn@doi [\mnras] {10.1093/mnras/sty1776}, \href
  {https://ui.adsabs.harvard.edu/abs/2018MNRAS.479.5544M} {479, 5544}

\bibitem[\protect\citeauthoryear{{Nally}}{{Nally}}{2023}]{starbug2}
{Nally} C.,  2023, {StarbugII: JWST PSF photometry for crowded fields},
  Astrophysics Source Code Library, record ascl:2309.012 (\mn@eprint {ascl}
  {2309.012})

\bibitem[\protect\citeauthoryear{{Nally} et~al.,}{{Nally}
  et~al.}{2024}]{Nally2024}
{Nally} C.,  et~al., 2024, \mn@doi [\mnras] {10.1093/mnras/stae1163}, \href
  {https://ui.adsabs.harvard.edu/abs/2024MNRAS.531..183N} {531, 183}

\bibitem[\protect\citeauthoryear{{Nayak} et~al.,}{{Nayak}
  et~al.}{2024}]{Nayak2024b}
{Nayak} O.,  et~al., 2024, \mn@doi [\apj] {10.3847/1538-4357/ad7baf}, \href
  {https://ui.adsabs.harvard.edu/abs/2024ApJ...975..262N} {975, 262}

\bibitem[\protect\citeauthoryear{{Neff}, {Eilek}  \& {Owen}}{{Neff}
  et~al.}{2015}]{Neff2015}
{Neff} S.~G.,  {Eilek} J.~A.,   {Owen} F.~N.,  2015, \mn@doi [\apj]
  {10.1088/0004-637X/802/2/87}, \href
  {https://ui.adsabs.harvard.edu/abs/2015ApJ...802...87N} {802, 87}

\bibitem[\protect\citeauthoryear{{Nesvadba} et~al.,}{{Nesvadba}
  et~al.}{2010}]{Nesvadba2010}
{Nesvadba} N.~P.~H.,  et~al., 2010, \mn@doi [\aap]
  {10.1051/0004-6361/200913333}, \href
  {https://ui.adsabs.harvard.edu/abs/2010A&A...521A..65N} {521, A65}

\bibitem[\protect\citeauthoryear{{Neumayer}, {Cappellari}, {Reunanen}, {Rix},
  {van der Werf}, {de Zeeuw}  \& {Davies}}{{Neumayer}
  et~al.}{2007}]{Neumayer2007}
{Neumayer} N.,  {Cappellari} M.,  {Reunanen} J.,  {Rix} H.-W.,  {van der Werf}
  P.~P.,  {de Zeeuw} P.~T.,   {Davies} R.~I.,  2007, \mn@doi [\apj]
  {10.1086/523039}, \href
  {https://ui.adsabs.harvard.edu/abs/2007ApJ...671.1329N} {671, 1329}

\bibitem[\protect\citeauthoryear{{Oosterloo} \& {Morganti}}{{Oosterloo} \&
  {Morganti}}{2005}]{Oosterloo2005}
{Oosterloo} T.~A.,  {Morganti} R.,  2005, \mn@doi [\aap]
  {10.1051/0004-6361:20041379}, \href
  {https://ui.adsabs.harvard.edu/abs/2005A&A...429..469O} {429, 469}

\bibitem[\protect\citeauthoryear{{Pantoni} et~al.,}{{Pantoni}
  et~al.}{2026}]{Pantoni2026}
{Pantoni} L.,  et~al., 2026, \mn@doi [\aap] {10.1051/0004-6361/202558839},
  \href {https://ui.adsabs.harvard.edu/abs/2026A&A...709A.237P} {709, A237}

\bibitem[\protect\citeauthoryear{{Peng}, {Ford}, {Freeman}  \& {White}}{{Peng}
  et~al.}{2002}]{Peng2002}
{Peng} E.~W.,  {Ford} H.~C.,  {Freeman} K.~C.,   {White} R.~L.,  2002, \mn@doi
  [\aj] {10.1086/344308}, \href
  {https://ui.adsabs.harvard.edu/abs/2002AJ....124.3144P} {124, 3144}

\bibitem[\protect\citeauthoryear{{Peng}, {Ford}  \& {Freeman}}{{Peng}
  et~al.}{2004}]{Peng2004b}
{Peng} E.~W.,  {Ford} H.~C.,   {Freeman} K.~C.,  2004, \mn@doi [\apj]
  {10.1086/381236}, \href
  {https://ui.adsabs.harvard.edu/abs/2004ApJ...602..705P} {602, 705}

\bibitem[\protect\citeauthoryear{{Perna} et~al.,}{{Perna}
  et~al.}{2020}]{Perna2020}
{Perna} M.,  et~al., 2020, \mn@doi [\aap] {10.1051/0004-6361/202038328}, \href
  {https://ui.adsabs.harvard.edu/abs/2020A&A...643A.139P} {643, A139}

\bibitem[\protect\citeauthoryear{{Quillen}, {de Zeeuw}, {Phinney}  \&
  {Phillips}}{{Quillen} et~al.}{1992}]{Quillen1992}
{Quillen} A.~C.,  {de Zeeuw} P.~T.,  {Phinney} E.~S.,   {Phillips} T.~G.,
  1992, \mn@doi [\apj] {10.1086/171329}, \href
  {https://ui.adsabs.harvard.edu/abs/1992ApJ...391..121Q} {391, 121}

\bibitem[\protect\citeauthoryear{{Quillen}, {Graham}  \& {Frogel}}{{Quillen}
  et~al.}{1993}]{Quillen1993}
{Quillen} A.~C.,  {Graham} J.~R.,   {Frogel} J.~A.,  1993, \mn@doi [\apj]
  {10.1086/172943}, \href
  {https://ui.adsabs.harvard.edu/abs/1993ApJ...412..550Q} {412, 550}

\bibitem[\protect\citeauthoryear{{Quillen}, {Brookes}, {Keene}, {Stern},
  {Lawrence}  \& {Werner}}{{Quillen} et~al.}{2006}]{Quillen2006}
{Quillen} A.~C.,  {Brookes} M.~H.,  {Keene} J.,  {Stern} D.,  {Lawrence} C.~R.,
    {Werner} M.~W.,  2006, \mn@doi [\apj] {10.1086/504418}, \href
  {https://ui.adsabs.harvard.edu/abs/2006ApJ...645.1092Q} {645, 1092}

\bibitem[\protect\citeauthoryear{{Quillen} et~al.,}{{Quillen}
  et~al.}{2008}]{Quillen2008}
{Quillen} A.~C.,  et~al., 2008, \mn@doi [\mnras]
  {10.1111/j.1365-2966.2007.12768.x}, \href
  {https://ui.adsabs.harvard.edu/abs/2008MNRAS.384.1469Q} {384, 1469}

\bibitem[\protect\citeauthoryear{{Quillen}, {Neumayer}, {Oosterloo}  \&
  {Espada}}{{Quillen} et~al.}{2010}]{Quillen2010}
{Quillen} A.~C.,  {Neumayer} N.,  {Oosterloo} T.,   {Espada} D.,  2010, \mn@doi
  [\pasa] {10.1071/AS09069}, \href
  {https://ui.adsabs.harvard.edu/abs/2010PASA...27..396Q} {27, 396}

\bibitem[\protect\citeauthoryear{{Reiter}, {Morse}, {Smith}, {Haworth}, {Kuhn}
  \& {Klaassen}}{{Reiter} et~al.}{2022}]{Reiter2022}
{Reiter} M.,  {Morse} J.~A.,  {Smith} N.,  {Haworth} T.~J.,  {Kuhn} M.~A.,
  {Klaassen} P.~D.,  2022, \mn@doi [\mnras] {10.1093/mnras/stac2820}, \href
  {https://ui.adsabs.harvard.edu/abs/2022MNRAS.517.5382R} {517, 5382}

\bibitem[\protect\citeauthoryear{{Rejkuba}, {Minniti}, {Silva}  \&
  {Bedding}}{{Rejkuba} et~al.}{2001}]{Rejkuba2001}
{Rejkuba} M.,  {Minniti} D.,  {Silva} D.~R.,   {Bedding} T.~R.,  2001, \mn@doi
  [\aap] {10.1051/0004-6361:20011315}, \href
  {https://ui.adsabs.harvard.edu/abs/2001A&A...379..781R} {379, 781}

\bibitem[\protect\citeauthoryear{{Rejkuba}, {Minniti}, {Silva}  \&
  {Bedding}}{{Rejkuba} et~al.}{2003}]{Rejkuba2003}
{Rejkuba} M.,  {Minniti} D.,  {Silva} D.~R.,   {Bedding} T.~R.,  2003, \mn@doi
  [\aap] {10.1051/0004-6361:20034056}, \href
  {https://ui.adsabs.harvard.edu/abs/2003A&A...411..351R} {411, 351}

\bibitem[\protect\citeauthoryear{{Rejkuba}, {Greggio}, {Harris}, {Harris}  \&
  {Peng}}{{Rejkuba} et~al.}{2005}]{Rejkuba2005}
{Rejkuba} M.,  {Greggio} L.,  {Harris} W.~E.,  {Harris} G. L.~H.,   {Peng}
  E.~W.,  2005, \mn@doi [\apj] {10.1086/432462}, \href
  {https://ui.adsabs.harvard.edu/abs/2005ApJ...631..262R} {631, 262}

\bibitem[\protect\citeauthoryear{{Rejkuba}, {Harris}, {Greggio},
  {Crnojevi{\'c}}  \& {Harris}}{{Rejkuba} et~al.}{2022}]{Rejkuba2022}
{Rejkuba} M.,  {Harris} W.~E.,  {Greggio} L.,  {Crnojevi{\'c}} D.,   {Harris}
  G.~L.~H.,  2022, \mn@doi [\aap] {10.1051/0004-6361/202141347}, \href
  {https://ui.adsabs.harvard.edu/abs/2022A&A...657A..41R} {657, A41}

\bibitem[\protect\citeauthoryear{{Riebel}, {Srinivasan}, {Sargent}  \&
  {Meixner}}{{Riebel} et~al.}{2012}]{Riebel2012}
{Riebel} D.,  {Srinivasan} S.,  {Sargent} B.,   {Meixner} M.,  2012, \mn@doi
  [\apj] {10.1088/0004-637X/753/1/71}, \href
  {https://ui.adsabs.harvard.edu/abs/2012ApJ...753...71R} {753, 71}

\bibitem[\protect\citeauthoryear{{Rieke} et~al.,}{{Rieke}
  et~al.}{2015}]{Rieke2015PASP}
{Rieke} G.~H.,  et~al., 2015, \mn@doi [\pasp] {10.1086/682252}, \href
  {https://ui.adsabs.harvard.edu/abs/2015PASP..127..584R} {127, 584}

\bibitem[\protect\citeauthoryear{{Rigby} et~al.,}{{Rigby}
  et~al.}{2023}]{Rigby2023}
{Rigby} J.,  et~al., 2023, \mn@doi [\pasp] {10.1088/1538-3873/acb293}, \href
  {https://ui.adsabs.harvard.edu/abs/2023PASP..135d8001R} {135, 048001}

\bibitem[\protect\citeauthoryear{{Rigopoulou} et~al.,}{{Rigopoulou}
  et~al.}{2024}]{Rigopoulou2024}
{Rigopoulou} D.,  et~al., 2024, \mn@doi [\mnras] {10.1093/mnras/stae1535},
  \href {https://ui.adsabs.harvard.edu/abs/2024MNRAS.532.1598R} {532, 1598}

\bibitem[\protect\citeauthoryear{{Robitaille}, {Whitney}, {Indebetouw}, {Wood}
  \& {Denzmore}}{{Robitaille} et~al.}{2006}]{Robitaille2006}
{Robitaille} T.~P.,  {Whitney} B.~A.,  {Indebetouw} R.,  {Wood} K.,
  {Denzmore} P.,  2006, \mn@doi [\apjs] {10.1086/508424}, \href
  {https://ui.adsabs.harvard.edu/abs/2006ApJS..167..256R} {167, 256}

\bibitem[\protect\citeauthoryear{{Salom{\'e}}, {Salom{\'e}},
  {Miville-Desch{\^e}nes}, {Combes}  \& {Hamer}}{{Salom{\'e}}
  et~al.}{2017}]{Salome2017}
{Salom{\'e}} Q.,  {Salom{\'e}} P.,  {Miville-Desch{\^e}nes} M.-A.,  {Combes}
  F.,   {Hamer} S.,  2017, \mn@doi [\aap] {10.1051/0004-6361/201731429}, \href
  {https://ui.adsabs.harvard.edu/abs/2017A&A...608A..98S} {608, A98}

\bibitem[\protect\citeauthoryear{{Santoro}, {Oonk}, {Morganti}  \&
  {Oosterloo}}{{Santoro} et~al.}{2015}]{Santoro2015}
{Santoro} F.,  {Oonk} J.~B.~R.,  {Morganti} R.,   {Oosterloo} T.,  2015,
  \mn@doi [\aap] {10.1051/0004-6361/201425103}, \href
  {https://ui.adsabs.harvard.edu/abs/2015A&A...574A..89S} {574, A89}

\bibitem[\protect\citeauthoryear{{Schiminovich}, {van Gorkom}, {van der Hulst}
  \& {Kasow}}{{Schiminovich} et~al.}{1994}]{Schiminovich1994}
{Schiminovich} D.,  {van Gorkom} J.~H.,  {van der Hulst} J.~M.,   {Kasow} S.,
  1994, \mn@doi [\apjl] {10.1086/187246}, \href
  {https://ui.adsabs.harvard.edu/abs/1994ApJ...423L.101S} {423, L101}

\bibitem[\protect\citeauthoryear{{Schreier}, {Capetti}, {Macchetto}, {Sparks}
  \& {Ford}}{{Schreier} et~al.}{1996}]{Schreier1996}
{Schreier} E.~J.,  {Capetti} A.,  {Macchetto} F.,  {Sparks} W.~B.,   {Ford}
  H.~J.,  1996, \mn@doi [\apj] {10.1086/176917}, \href
  {https://ui.adsabs.harvard.edu/abs/1996ApJ...459..535S} {459, 535}

\bibitem[\protect\citeauthoryear{{Seale} et~al.,}{{Seale}
  et~al.}{2014}]{Seale2014}
{Seale} J.~P.,  et~al., 2014, \mn@doi [\aj] {10.1088/0004-6256/148/6/124},
  \href {https://ui.adsabs.harvard.edu/abs/2014AJ....148..124S} {148, 124}

\bibitem[\protect\citeauthoryear{{Shin}, {Woo}, {Chung}, {Baek}, {Cho}, {Kang}
  \& {Bae}}{{Shin} et~al.}{2019}]{Shin2019}
{Shin} J.,  {Woo} J.-H.,  {Chung} A.,  {Baek} J.,  {Cho} K.,  {Kang} D.,
  {Bae} H.-J.,  2019, \mn@doi [\apj] {10.3847/1538-4357/ab2e72}, \href
  {https://ui.adsabs.harvard.edu/abs/2019ApJ...881..147S} {881, 147}

\bibitem[\protect\citeauthoryear{{Sloan} et~al.,}{{Sloan}
  et~al.}{2007}]{Sloan2007}
{Sloan} G.~C.,  et~al., 2007, \mn@doi [\apj] {10.1086/519236}, \href
  {https://ui.adsabs.harvard.edu/abs/2007ApJ...664.1144S} {664, 1144}

\bibitem[\protect\citeauthoryear{{Soria} et~al.,}{{Soria}
  et~al.}{1996}]{Soria1996}
{Soria} R.,  et~al., 1996, \mn@doi [\apj] {10.1086/177403}, \href
  {https://ui.adsabs.harvard.edu/abs/1996ApJ...465...79S} {465, 79}

\bibitem[\protect\citeauthoryear{{Srinivasan} et~al.,}{{Srinivasan}
  et~al.}{2009}]{Srinivasan2009}
{Srinivasan} S.,  et~al., 2009, \mn@doi [\aj] {10.1088/0004-6256/137/6/4810},
  \href {https://ui.adsabs.harvard.edu/abs/2009AJ....137.4810S} {137, 4810}

\bibitem[\protect\citeauthoryear{{Struve}, {Oosterloo}, {Morganti}  \&
  {Saripalli}}{{Struve} et~al.}{2010}]{Struve2010}
{Struve} C.,  {Oosterloo} T.~A.,  {Morganti} R.,   {Saripalli} L.,  2010,
  \mn@doi [\aap] {10.1051/0004-6361/201014355}, \href
  {https://ui.adsabs.harvard.edu/abs/2010A&A...515A..67S} {515, A67}

\bibitem[\protect\citeauthoryear{{Talbot}, {Sijacki}  \& {Bourne}}{{Talbot}
  et~al.}{2022}]{Talbot2022}
{Talbot} R.~Y.,  {Sijacki} D.,   {Bourne} M.~A.,  2022, \mn@doi [\mnras]
  {10.1093/mnras/stac1566}, \href
  {https://ui.adsabs.harvard.edu/abs/2022MNRAS.514.4535T} {514, 4535}

\bibitem[\protect\citeauthoryear{{Taylor}}{{Taylor}}{2005}]{Taylor2005}
{Taylor} M.~B.,  2005, in {Shopbell} P.,  {Britton} M.,   {Ebert} R.,  eds,
  Astronomical Society of the Pacific Conference Series Vol. 347, Astronomical
  Data Analysis Software and Systems XIV. p.~29

\bibitem[\protect\citeauthoryear{{Villanueva} et~al.,}{{Villanueva}
  et~al.}{2025}]{Villanueva2025}
{Villanueva} V.,  et~al., 2025, \mn@doi [\aap] {10.1051/0004-6361/202553891},
  \href {https://ui.adsabs.harvard.edu/abs/2025A&A...695A.202V} {695, A202}

\bibitem[\protect\citeauthoryear{{Wang}, {Hammer}, {Rejkuba}, {Crnojevi{\'c}}
  \& {Yang}}{{Wang} et~al.}{2020}]{Wang2020}
{Wang} J.,  {Hammer} F.,  {Rejkuba} M.,  {Crnojevi{\'c}} D.,   {Yang} Y.,
  2020, \mn@doi [\mnras] {10.1093/mnras/staa2508}, \href
  {https://ui.adsabs.harvard.edu/abs/2020MNRAS.498.2766W} {498, 2766}

\bibitem[\protect\citeauthoryear{{Williams}, {van der Hucht}  \&
  {The}}{{Williams} et~al.}{1987}]{Williams1987}
{Williams} P.~M.,  {van der Hucht} K.~A.,   {The} P.~S.,  1987, \aap, \href
  {https://ui.adsabs.harvard.edu/abs/1987A&A...182...91W} {182, 91}

\bibitem[\protect\citeauthoryear{{Wright} et~al.,}{{Wright}
  et~al.}{2023}]{Wright2023}
{Wright} G.~S.,  et~al., 2023, \mn@doi [\pasp] {10.1088/1538-3873/acbe66},
  \href {https://ui.adsabs.harvard.edu/abs/2023PASP..135d8003W} {135, 048003}

\bibitem[\protect\citeauthoryear{{Wykes} et~al.,}{{Wykes}
  et~al.}{2013}]{Wykes2013}
{Wykes} S.,  et~al., 2013, \mn@doi [\aap] {10.1051/0004-6361/201321622}, \href
  {https://ui.adsabs.harvard.edu/abs/2013A&A...558A..19W} {558, A19}

\bibitem[\protect\citeauthoryear{{Yasui}, {Izumi}, {Saito}, {Lau}, {Kobayashi}
  \& {Ressler}}{{Yasui} et~al.}{2026}]{Yasui2026}
{Yasui} C.,  {Izumi} N.,  {Saito} M.,  {Lau} R.~M.,  {Kobayashi} N.,
  {Ressler} M.~E.,  2026, \mn@doi [\apj] {10.3847/1538-4357/ae56fe}, \href
  {https://ui.adsabs.harvard.edu/abs/2026ApJ..1002...73Y} {1002, 73}

\bibitem[\protect\citeauthoryear{{Zeidler}, {Sabbi}, {Nota}, {Manjavacas},
  {Jones}  \& {Pacifici}}{{Zeidler} et~al.}{2024}]{Zeidler2024}
{Zeidler} P.,  {Sabbi} E.,  {Nota} A.,  {Manjavacas} E.,  {Jones} O.~C.,
  {Pacifici} C.,  2024, \mn@doi [\apj] {10.3847/1538-4357/ad779e}, \href
  {https://ui.adsabs.harvard.edu/abs/2024ApJ...975...18Z} {975, 18}

\bibitem[\protect\citeauthoryear{{van Gorkom}, {van der Hulst}, {Haschick}  \&
  {Tubbs}}{{van Gorkom} et~al.}{1990}]{vanGorkom1990}
{van Gorkom} J.~H.,  {van der Hulst} J.~M.,  {Haschick} A.~D.,   {Tubbs} A.~D.,
   1990, \mn@doi [\aj] {10.1086/115456}, \href
  {https://ui.adsabs.harvard.edu/abs/1990AJ.....99.1781V} {99, 1781}

\bibitem[\protect\citeauthoryear{{van den Bergh}}{{van den
  Bergh}}{1976}]{vandenBergh1976}
{van den Bergh} S.,  1976, \mn@doi [\apj] {10.1086/154648}, \href
  {https://ui.adsabs.harvard.edu/abs/1976ApJ...208..673V} {208, 673}

\makeatother
\end{thebibliography}

\input{main.bbl} 







\bsp	
\label{lastpage}
\end{document}